\documentstyle[aps,preprint,eqsecnum]{revtex} 
\def\alp#1{\mbox{$\alpha_ #1 ^{-1}$}}
\def\msu{\mbox{$\overline{MS}$}}
\def\sf{{\scriptscriptstyle F}}
\def\sg{{\scriptscriptstyle G}}

\def\ss{{\scriptscriptstyle S}}

\def\sw{{\scriptscriptstyle W}}
\def\sz{{\scriptscriptstyle Z}}
\def\sgam{{\scriptscriptstyle \Gamma}}
\def\mz{\mbox{$m_{Z}$}}
\def\mzz{\mbox{$m_{Z}^2$}}
\def\mw{\mbox{$m_{W}$}}
\def\mt{\mbox{$m_{t}$}}
\def\mh{\mbox{$m_{h}$}}
\def\mfi{\mbox{$M_{\Phi}$}}
\def\msi{\mbox{$M_{\Sigma}$}}
\def\msw{\mbox{$m_{1/2}$}}
\def\sio{\mbox{$\xi_0$}}
\def\m#1{\mbox{$m_{\tilde{#1}}$}}

\def\ms#1{\mbox{$m_{\tilde{#1}}^2$}}

\def\al#1#2{\mbox{$\alpha_#1^{-1}\mid_{#2}$}}
\def\alzi{\mbox{$\alpha_3^{-1}(\mz)$}}

\def\alz{\mbox{$\alpha_3(m_Z)$}}
\def\plb#1#2#3{ Phys. Lett. B #1 (#2) #3}

\def\npb#1#2#3{ Nucl. Phys. B #1 (#2) #3}

\def\prd#1#2#3{ Phys. Rev.  D #1 (#2) #3}
\def\pr#1#2#3{ Phys. Rev.   #1 (#2) #3}
\def\prl#1#2#3{ Phys. Rev. Lett.  #1 (#2) #3}
\def\zpc#1#2#3{ Z Phys. C #1 (#2) #3}
\def\nca#1#2#3{ Nuovo Cimento A #1 (#2) #3}
\tighten
\begin{document}
\draft

\preprint{ LAEFF--95--01}
\title
{
 Effective couplings and Perturbative  Unification
}
\author{M. Bastero--Gil\cite{aa1} and J.P\'erez--Mercader\cite{aa2}}
\address{
Laboratorio de Astrof\'{\i}sica Espacial y F\'{\i}sica
Fundamental\\
Apartado 50727\\
28080 Madrid
}
\maketitle
\begin{abstract}

In this paper we study the influence of the threshold effects due to
massive degrees of freedom in the evolution with scale of gauge
coupling constants. We first describe in detail the (standard) mass
dependent renormalization prescription we use. This guides us to
introduce and work with effective couplings, which are finite, process
independent, and include complete threshold effects. We compute the
evolution of the effective couplings in both, the Standard Model and
its Minimal Supersymmetric extension, from \mz\ to the high energy
region. We find that the effects from thresholds due to the standard
massive gauge bosons are non--negligible, contrary to what is
generally assumed when using other, less--accurate, descriptions of
the thresholds, as for example in the step--function
approximation. Moreover, we find that thresholds are relevant when
studying perturbative SUSY unification, changing the conclusions
reached when using the step--function approach. We find that threshold
effects bring conflict between the known experimental data at $m_Z$,
the naturalness upper bound on the masses of SUSY partners and the
$perturbative$ unification of couplings.

\end{abstract}
\vspace{15mm}
\begin{center}
{\it To be published in Nuclear Physics B}
\end{center}
\pagebreak
\section{Introduction.}
\label{sect1}
In studying Grand Unification theories for the strong, weak and
electromagnetig interactions, we are compelled to work with two
different scales: the electroweak symmetry breaking scale ($O(m_Z)$),
on one hand, and the scale of the symmetry breaking of the unifying
group $G$, or unification scale ($M_X \gtrsim 10^{16}\,GeV$), on the
other. In general, the unification schemes assume the existence of a
``desert" between these scales (separated, at least, by 13 orders of
magnitude), so that no new effects are present, beyond those coming
from the unification group $G$.

To test the validity of the unification picture, one studies the
evolution of the three gauge couplings of the Standard Model, from low
($m_Z$) to high energies ($M_X$), seeking for their convergence
towards a common value at and beyond the unification scale $M_X$. In
the simplest scenario, we can neglect the effects of the heavy degrees
of freedom introduced by $G$ in the evolution of the running
couplings, and work only with the content of matter of the low energy
theory (light degrees of freedom). This would be a first step in
towards checking unification for the model. If, in this approximation,
the couplings meet at a common value, $g$, at some point of the high
energy region, $M_X$, we will have given the first step in the right
direction.

The following step would consist of including the effects of the heavy
degrees of freedom in our scheme. At low energy, these degrees of
freedom are decoupled from the theory, but they become operative as
soon as we approach the scale $M_X$, changing smoothly the evolution
of the couplings. To take into account their effects in a proper way,
one can integrate out the heavy fields from the complete action of the
full theory, $S[G]$ \cite{lam2,lam1}. Carrying this out, one obtains
as unification condition:
$g_i^{-2}(\ms)=g^{-2}(\mu)+\lambda_i(\mu,M_j)$, where the functions
$\lambda_i(\mu,M_j)$ depend logarithmically on the masses of the heavy
fields $M_j \approx O(M_X)$, the scale $\mu$ satisfies the condition
$m_i \ll \mu \ll M_j$, with $m_i$ the masses of the light fields. With
this construction, the behavior of $g_i(\mu)$ is only dictated by the
light degrees of freedom.

 The main tool to carry out this two--step program are the
renormalization group equations (RGE) and in particular the $\beta$--
functions for the coupling constants, defined as
$\displaystyle{\beta_i=\frac{d g_i(\mu)}{d \ln \mu}}$. In most of the
Grand Unification analyses carried out, the $\beta$--functions used
are those computed using the modified minimal subtraction
($\overline{MS}$) scheme \cite{msm}. But, since there is no unique
renormalization prescription to work with, and different prescriptions
yield different explicit forms of the $\beta$-- functions, the
question that naturally arises is {\it to which extent the results on
unification depend on the prescription chosen}, that is, on the way
one studies the evolution of the coupling constants with scale.

 The choice of renormalization prescription is not a matter of
taste. It implies the way one deals with physical effects, such as
threshold effects, coming from the presence of massive particles in
the theory. The \msu\ procedure is only one of the possible choices of
a generic class of subtraction procedures, called mass independent
subtraction procedures (MISP) \cite{misp}.  For this class, the
$\beta$--functions depend only on the particle $content$ at the energy
scale at which one computes them, and not explicitly on the masses of
the particles. On the other hand, there also is available another
generic class of procedures, called mass dependent subtraction
procedures (MDSP) \cite{mdsp}, which take into account the dependence
{\it not only} on the number, {\it but also} on the particle
$masses$. While with the resulting MISP--$\beta$--functions one is
forced to put in by hand the information about the mass spectrum of
the theory when integrating the RGEs, with MDSPs one includes, in a
natural way, all the information about threshold effects and possible
decoupling of the massive particles in the $evolution$ of the coupling
constant \cite{desac}.  Because of this feature, the MDSPs are
threfore more complete and precise than the MISPs. They have the
disadvantage that the calculation of the $\beta$--functions in MDSPs
is far more complicated from a technical point of view.

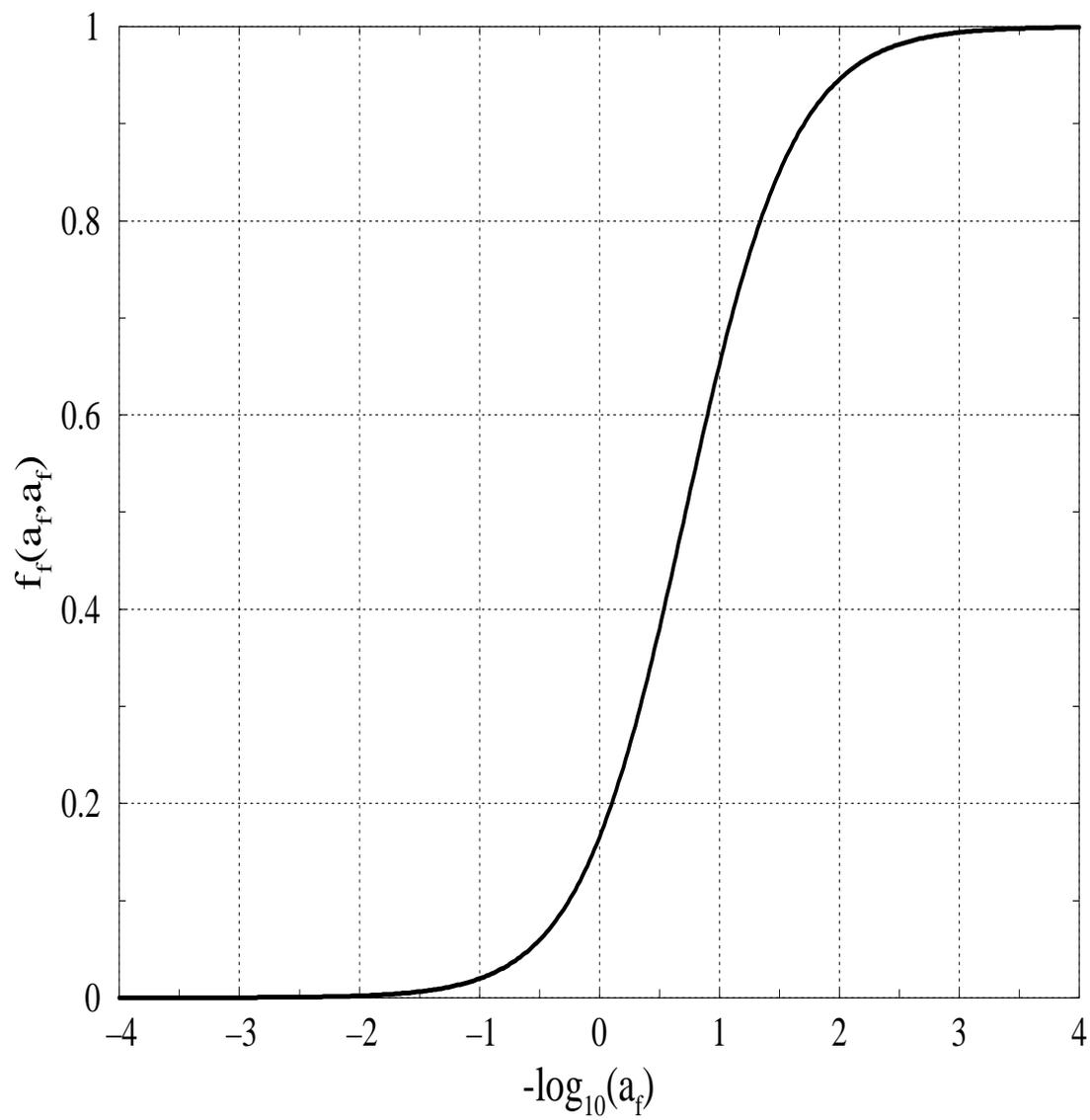
\begin{figure}\caption{Function $f_f(a_f)$ respect to
$-log_{10}(a_f)=log_{10}(q^2/m_f^2)$. As
it is shown in the plot, the function tends to 1 in the limit of neglecting
mass respect to $q^2$ ($-log_{10}(a_f) \rightarrow \infty$),
while it tends to 0 (decoupling) in the limit of heavy mass
($-log_{10}(a_f) \rightarrow -\infty$). \label{derivf2}}
\end{figure}

Working with the \msu\ $\beta$--functions, the standard procedure to
take into account the contribution of a particle of a given mass $m$
at same scale $\mu$, consists in making use of a ``step--function''
$\theta (\mu^2-m^2)$: one for each massive particle \cite{marciano}.
This method constrains the contribution of a particle strictly to
scales higher than its mass, being zero otherwise, and gives only the
dominant logarithmic contribution of the mass. On the contrary, in the
MDSP--$\beta$--function, the contribution of each massive degree of
freedom is controlled via a $smooth$ function $f(m/\mu)$, calculated
in perturbation theory, that has the limits $f(m/\mu) \rightarrow 1$
when the ratio $(m/\mu)$ goes to zero, and $f(m/\mu) \rightarrow 0$
when $(m/\mu)$ goes to infinity.  They give a non--zero contribution
even for scales $\mu \leq m$, and a more accurate description of the
threshold crossing than the step--function approximation (see for
example, Fig. (\ref{derivf2})).  This behavior is a
quantum--mechanical effect, reflecting the Heisenberg uncertainty
principle: we have a non--zero probability of producing a particle
even for momenta below the mass scale of the particle; furthermore,
the contribution of the degree of freedom to the $\beta$--function
spreads over a few orders of magnitude in momentum.

On the different treatment of threshold crossing resides the main
source for the discrepancies in the final results derived with one or
another approach. The change in the derivative of the coupling
constant when crossing the threshold will affect the value of the
couplings even when we are away from the threshold region at several
orders of magnitude above the particle mass scale. And this will have
clear implications in the study of unification theories with a rich
low energy mass spectrum, such as for the Minimal Supersymmetric
extension of the Standard Model (MSSM).

Previous work on the unification of the minimal susy model indicates
that unification of gauge couplings is possible, and compatible with a
susy spectrum below $1\;TeV$ and a unification scale of
$O(10^{16}\;GeV)$ \cite{amaldi,zichichi}.  These results were carried
out by using the \msu\ $\beta$--functions, corrected with
step--functions to treat the susy spectrum. Here, we will instead work
with a different renormalization prescription, a mass dependent
prescription, to include threshold effects.  We will compare the
results with those obtained by other authors. In this way, we will try
to understand where and why are the differences between the
conclusions reached with different approaches.

With the derivation we use to get threshold effects, we aim to clarify
the point that these effects are not higher order corrections with
respect the order of perturbation theory one is working, but instead
they have to be calculated at $each$ order of perturbation theory.

Before dealing with the susy model, we need, of course, to fix our
prescription. We would have to choose one of the possible
MDSPs. However, better than pick up a particular procedure and compute
the $\beta$--function, we will see that the use of an MDSP allows us
to define effective couplings, analogous to the effective charges
defined by Gell--Mann and Low, Stuckelberg and Peterman in QED
\cite{gell}, in the sense that they are finite, universal, and include
complete threshold effects, being by the way appropriate for our
purposes. This will be the subject of Section 2. We will end that
section with the definition of gauge effective couplings associated
with the gauge group $SU(3)_c \times SU(2)_L \times U(1)_Y$. Some
problems related with the definition of the effective couplings for
non--abelian theories, and also with broken symmetries, are also
discussed in Section 2.

In Section 3 we first study the evolution of the gauge couplings of
the SM with minimal matter contents: three generations of fermions,
and one doublet of scalars. In this case, the number of massive
degrees of freedom in the range ($m_Z$, $M_X$) is small, and we do not
expect the unification of couplings to take place. In spite of this,
the model becomes useful to point out the importance of threshold
effects associated with massive gauge bosons, $W^{\pm}$ and $Z^0$,
otherwise neglected in the step--function approximation.

We should keep in mind these contributions when studying the minimal
susy extension of the SM, where more degrees of freedom (susy
partners) are present beyond the scale $m_Z$. We will see in Section 4
that gauge boson threshold effects become relevant in attaining or not
the ``unification'' of the gauge couplings. We first examine the model
in the simplest approach (one--loop without unification gauge group
thresholds); after that we will consider corrections due to heavy
threshold effects in the standard way described in the beginning of
this Introduction, and also will discuss the effects due to two--loop
corrections.

In Section 5 we offer our conclusions, and discuss other approaches
that have emerged in the recent literature
\cite{lynn,grinstein,bagger}.  The works cited on Ref. \cite{bagger}
remark the importance of complete light (susy) threshold effects for
the determination of the \msu\ values of the gauge couplings at the
electroweak scale, as they are extracted from experimental quantities,
and the later implication of these initial values for unification.
Susy thresholds distinguish between the quantities extracted using the
SM, from the same quantities obtained on the MSSM. The main point of
these papers is that not only the leading (logarithmic) threshold
contributions are important, but also the non--leading, non
logarithmic terms. In some sense, that is also our main statement. But
we remark that here we study the impact of thresholds in the
$evolution$ with scale. Since we aim at comparing different
evolutions, as given by two different approaches for the thresholds,
we will assume the same initial conditions for the couplings in both
schemes\footnote{Further work in the determination of the initial
values for the effective couplings at the electroweak scale in our
scheme, is now in progress.}.

\section{Effective Couplings.}
\label{sect2}
  Our purpose in this Section is to define effective couplings,
following the simplest possible arguments from the general theory of
renormalization. This is an easy task when the coupling is the
coupling constant for an abelian gauge theory, as in QED. We will use
what we learn in this case as a guide to get at the end the expression
for non--abelian effective couplings.

For a general theory, the renormalized coupling, $g_r$, is related to
the bare coupling, $g_0$, by $g_r=Zg_0$, where $Z$ is a product of
renormalization constants: $Z=Z^{1/2}_3 (Z_2 Z_1^{-1})_a$. Here, $Z_3$
is the wave function renormalization constant (WFRC) of the gauge
boson which defines the interaction; $Z_2$ is the product of WFRC of
the external legs; and $Z_1$ is the proper vertex renormalization
constant. The subindex ``$a$" refers to the different classes of
vertices involving the same coupling constant, which come from the
interaction of the gauge boson with itself or with the other particles
present in the model: fermions, scalars, $\ldots$. When using a MISP,
the Slavnov--Taylor identities (Ward--Takahashi for abelian theories)
guarantee that $(Z_2 Z_1^{-1})_a$ is independent of vertex we choose
to define the coupling, and thus we obtain in this case a $universal$
renormalized coupling, albeit without including threshold
effects. With a MDSP, we include the dependence on the masses, in
particular on the masses of the external legs, and we could, in
principle, distinguish among the renormalized couplings derived from
different processes: we would get a mass--dependent coupling, but {\it
no a universal} coupling \cite{ross}. This would not be inconsistent
with the constraints imposed by the Slavnov--Taylor identities. For
example, one can define the renormalization constants associated with
a specific process, and calculate the $\beta$--function, and after
make use of the Slavnov--Taylor identities to derive the remaining
renormalization constants for the theory, without modifying the
previously defined $\beta$--function\footnote{This was explicitly
shown for QCD in Ref. \cite{bc3}.}. But the renormalized coupling
would not be universal (independent of the process). Therefore, in
order to get a universal renormalized coupling, we first have to make
sure that the MDSP respects the Slavnov--Taylor identities, using them
to select the finite contributions (which contain the threshold
effects) to the $Z_i$.  This will guarantee the gauge invariance
of our effective couplings.

For abelian gauge theories, such as QED, one can extend the
Ward--Takahashi identity, $Z_2=Z_1$, to the finite terms of the $Z_i$,
and therefore the relation $e_r=Z_3^{1/2}g_0$ is valid for mass
independent and mass dependent procedures. Moreover, we can see that,
if we choose a suitable mass dependent procedure, the renormalized
charge is equivalent to the effective charge of Gell--Mann and Low,
Stuckelberg and Peterman, defined by:
\begin{equation}
e^2_{eff}(q^2)=\frac{e^2_0}{1+\Pi^T_0(q^2)}\,. \label{eeff}
\end{equation}
Here, $\Pi^T_0(q^2)$ is the $transverse$ component of the bare vacuum
polarization tensor of the gauge boson. The effective charge is (a)
independent of the renormalization scheme, since it is defined via
bare functions, (b) universal and gauge independent and, (c) finite.

 The last point is easily demostrated when one replaces the bare
coupling by the renormalized coupling in (\ref{eeff}), since $Z_3$
must verify that the product $Z_3(1+\Pi^T_0(q^2))$ be a finite
function of both the scale $q$ and the subtraction point $\mu$.  We
therefore have:
\begin{equation}
e^2_{eff}(q^2)=\frac{e^2_r(\mu)}{(1+\Pi^T_r(q^2,\mu^2))}\,,
\end{equation}
 where the explicit form of $\Pi^T_r(q^2,\mu^2)$ will depend on
the chosen subtraction procedure.

In general, the two couplings $e^2_{eff}$ and $e^2_r$ are not
equivalent. For example, the renormalized charge using $\overline{MS}$
does not include threshold effects due to massive fermions, whereas
they are present in $e^2_{eff}$ through the functions
$\Pi^T_r(q^2,\mu^2)$. On the contrary, if we adopt a mass dependent
procedure with the normalization condition $1+\Pi^T_r(\mu^2,\mu^2)=1$,
then $e^2_{eff}(\mu^2)=e^2_r(\mu^2)$, and
$$
\frac{\beta(e_{eff}^2)}{e_{eff}^2}=\frac{\beta(e_{r}^2)}{e_{r}^2}\,.
$$
Therefore, both constants depend on the scale in the same way, and
they include the threshold effects.

In the case of non--abelian theories, to get a finite expression for
the effective couplings it is necessary something beyond the
correction to the vacuum polarization. We must supply another
function, $\Gamma_0(q^2)$, to the definition (\ref{eeff}),
\begin{equation}
g^2_{eff}(q^2)=\frac{g^2_0}{(1+\Pi^T_0(q^2,\mu^2))(1+2
\Gamma_0(q^2))}= \frac{g^2_r(\mu)}{Z_3(1+\Pi^T_0(q^2,\mu^2))
(Z_2Z_1^{-1})_a^2 (1+2 \Gamma_0(q^2))}\,,
\end{equation}
which comes from the corrections due to vertex and external legs, and
which must verify that the product $(Z_2Z_1^{-1})_a(1+\Gamma_0(q^2))$
be a finite function. The function $\Gamma_0(q^2)$ consists of a
divergent term, $\Gamma_0^{div}$, and a finite contribution which
includes a dependence on the masses. The Slavnov--Taylor identities
guarantee that the divergent term of $(Z_2 Z_1^{-1})_a$ is the same
for all possible vertices ``$a$", and is given by \cite{zg}:
\begin{equation}
(Z_2Z_1^{-1})_a^{div}=1+\frac{g^2}{(4\pi)^2}C_2(G)\frac{3+\xi}{4}
\left(\frac{2}{n-4}\right)\,, \label{z21div}
\end{equation}
where $C_2(G)$ is the quadratic Casimir of the non--abelian group $G$,
$\xi$ is the gauge parameter, and $n$ is the dimension of the
space--time. Thus, $1-\Gamma_0^{div}=(Z_2Z_1^{-1})_a^{div}$.

But, it is not straightforward to calculate the finite contributions
to $\Gamma_0(q^2)$, {\it i.e.}, to define $(Z_2Z_1^{-1})_a$, with a
mass--dependent procedure. The problem is to make sure that the
resulting $\Gamma_0(q^2)$ is independent of the vertex ``$a$" chosen
to perform the calculation. For example, if we choose the vertex with
fermions on the external legs, we can expect that $(Z_2Z_1^{-1})_f$
depends on the fermion masses, the scalar masses, and the gauge boson
masses (if the symmetry is broken). The same kind of masses may appear
if we consider the scalar--gauge boson vertex, or the trilinear boson
vertex. However, if we choose the ghost--gauge boson vertex, we only
have running in the loop, ghosts, gauge bosons or Goldstone bosons (if
the symmetry is broken): either massless particles, for unbroken
symmetry, or particles with masses proportional to the gauge boson
mass, for broken symmetry. Now, if we impose the Slavnov--Taylor
identities, and thus $(Z_2Z_1^{-1})_f=(Z_2Z_1^{-1})_{ghost}=\ldots$,
it is clear that the $universal$ correction $\Gamma_0(q^2)$ can not
depend on the fermion or scalar masses, but only {\it on the gauge
boson masses}.

In this sense, the universal threshold effects associated with the
vertex are related to processes leading to gauge bosons production.
Therefore, if we have an unbroken symmetry, there are no threshold
effects from the vertex in the effective couplings, because we can
always produce a massless particle. The corrections due to the
presence in the vertex of other massive particles, such as fermions or
scalars, will affect other parameters of the theory, but not the
effective couplings. Moreover, these process depending corrections are
finite, as it is implied by the Slavnov--Taylor identities.

So far, we have discussed about the general expressions of the
effective couplings (for a general non abelian theory) given in terms
of the functions $\Pi_0^T(q^2)$ and $\Gamma_0(q^2)$. The transverse
bare vacuum polarization is easily calculated from the appropriate
Feynman diagrams.  We know the divergent part and the masses presents
in the finite term of $\Gamma_0(q^2)$, but the explicit form of
$\Gamma_0(q^2)$ will be closely related to the specific gauge theory
we treat.

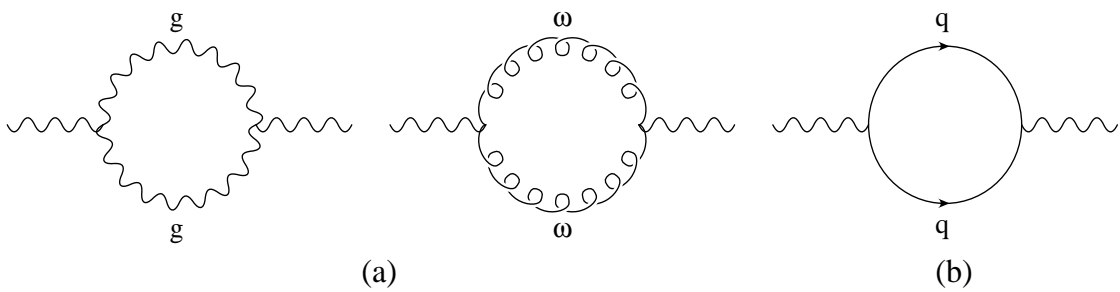
\begin{figure}\caption{ The 1--loop particle contributions to the $\Pi^T_{gg}$
function. (a) Gluon contributions plus ghost
($\omega_i$) contributions. (b) Fermion contributions. \label{figgluon}}
\end{figure}

In particular, we are interested in studying the evolution of the
gauge couplings of the Standard Model, $g_3$, $g$ and $g'$ of $SU(3)_c
\times SU(2)_L \times U(1)_Y$. We have no problems with the definition
of $g_3(q^2)$, since QCD is an exact non abelian theory, so that:
\begin{equation}
g_3(q^2)=\frac{g_3^2}{1+\Pi_{gg}^T(q^2)+2 \Gamma_3(q^2)}\,,
\end{equation}
where,
\begin{equation}
\Gamma_3(q^2)=-g_3^2\frac{3}{4}(3+\xi)\left(
\frac{2}{n-4}-\ln\frac{q^2}{\mu^2}+Constant\right) \,,
\end{equation}
and $\Pi_{gg}^T(q^2)$ is the transverse bare vacuum polarization of
the gluon, given by the diagrams of Fig. (\ref{figgluon}).

 To define the other two effective couplings, $g$ and $g'$, we
have to take into account that $SU(2)_L \times U(1)_Y$ is a broken
symmetry at low energy (electroweak scale), just where we begin to run
the couplings. In the broken phase, we have the three gauge bosons
(the eigenstates of the mass matrix) $W^{\pm}$, $Z^0$ and the photon,
$A$.  We can define the effective coupling $g^2(q^2)$ by the
interaction of the $W^{\pm}$; however, since in this phase the gauge
boson $B$, associated with the gauge symmetry $U(1)_Y$, is a mixed
state of the neutral gauge bosons, it is better to work with the
electromagnetic coupling, $e^2(q^2)$, given by the interaction of the
photon, and define $g'$ through the equation:
\begin{equation}
\frac{1}{g'^2}=\frac{1}{e^2}-\frac{1}{g^2}\,. \label{prima}
\end{equation}

In the Standard Model, $e^2(q^2)$ is not a pure abelian coupling, so
that to define $g(q^2)$ and also $e^2(q^2)$ we need the correction
$\Gamma(q^2,m^2)$, where now ``$m$" may be $m_Z$ or $m_W$. This vertex
correction is the same for both couplings, because the non abelian
character of $e^2$ is closely related to $g^2$
($e^{-2}=g^{-2}+g'^{-2}$). To obtain its explicit form we follow the
argument given by Kennedy and Lynn \cite{kl}, and that we partially
reproduced in the following.

 These authors make use of the relation between the non--abelian
vertex correction, $\Gamma(q^2)$, and the longitudinal term of the
mixed vacuum polarization tensor for the neutral bosons,
$\Pi_{ZA}^L(q^2)$ \cite{sirlin}. The latter contribution gives rise to
a non--diagonal mass matrix for the $(Z^0,\,A)$ system.  Therefore,
one needs to redefine the fields $Z^0$ and $A$, which are the correct
mass eigenstates at tree--level, in order to eliminate the
non--diagonal term, and get the correct eigenstates (and a massless
photon!) at one--loop order.

This can be carried out by first redefining the coupling $g$,
including the universal vertex correction $\Gamma$ with
$\tilde{g}=g(1-\Gamma)$, and then defining new fields $\tilde{Z}$ and
$\tilde{A}$ making use of this coupling. Now, in this basis the
non-diagonal term of the mass matrix becomes $\Pi_{ZA}^L+ m^2_Z g g'
\Gamma /(g^2+g'^2)$. Thus, if we choose $\Gamma$ to satisfy the
condition
\begin{equation}
\Pi_{ZA}^L+m^2_Z \frac{gg'}{g^2+g'^2}\Gamma=0 \,, \label{defgam1}
\end{equation}
we get the desired results that: (a) the photon remains massless (at
least at one--loop), and (b) we determine the explicit form of
$\Gamma(q^2)$.  Obviously, condition (\ref{defgam1}) is an identity
for the divergent terms of $\Pi_{ZA}^L$ and $\Gamma$. What we get with
(\ref{defgam1}) is the finite and mass--dependent term of $\Gamma$.

Now, we have all the ingredients involved in the definitions of
$g^2(q^2)$ and $e^2(q^2)$, which are given by:
\begin{equation}
g^2(q^2)=\frac{g^2}{1+\Pi_{WW}^T(q^2)+2 \Gamma (q^2)}\;,\;\;
e^2(q^2)=\frac{e)^2}{1+\Pi_{AA}^T(q^2)+2 \frac{e^2}{g^2} \Gamma
(q^2)}\,,
\end{equation}
where $\Pi_{WW}^T$, $\Pi_{AA}^T$ are the transverse bare vacuum
polarization tensor of the $W^{\pm}$ and the photon respectively.

In Appendix A we give the expressions of the functions $\Pi_{WW}^T$,
$\Pi_{AA}^T$ and $\Gamma(q^2)$, as well as $\Pi_{gg}^T$. All these
functions, including $\Gamma_3(q^2)$, depend on the gauge parameter
$\xi$. This dependence cancels exactly for the divergent part of the
combinations $\Pi_{ii}^T+\Gamma_i$, but the same does not happen for
the finite term due to the presence of degrees of freedom whose masses
are proportional to $\xi$ (i.e., gauge bosons, ghosts, and except in
$\Pi_{gg}^T$ and $\Gamma_3(q^2)$, also Goldstone bosons). This leads
to effective couplings depending on the gauge parameter \cite{bc3}.

The gauge parameter also changes with scale, as given by its RGE
\begin{equation}
\frac{d \xi}{d \ln q^2}= -\xi \frac{d \ln Z_3}{d \ln q^2} \,,
\end{equation}
and it will depend on the associated\footnote{We will have four
different gauge parameters, associated with the bosons $W^{\pm}$, $A$
and $Z^0$ and gluons respectively.} coupling $g_i^2(q^2)$ through the
dependence of $Z_3$. In order to get the evolution of the effective
couplings we have to solve a system of coupled differential
equations. The change in $g_i^2(q^2)$ produced by the change in
$\xi(q^2)$ will be small (order two--loop); and if we calculate the
effective coupling at one--loop order, we can neglect it, and maintain
$\xi(q^2)$ at its initial value. But if we calculate $g_i^2(q^2)$ at
two--loop order we need, at least, $\xi(q^2)$ at one--loop order, and
so on. On the other hand, the theory by itself indicates the most
suitable value of $\xi$ to work with, without approximations: the
fixed point of the differential equation at $\xi=0$. Thus, we choose
to work in the Landau gauge.

Up to now, we have defined the effective couplings in the Standard
Model, which are {\it finite, universal and include threshold
effects}. For the $SU(2)_L \times U(1)_Y$ couplings, we made partially
use of the arguments given by Kennedy and Lynn to derive the function
$\Gamma(q^2)$. Nevertheless, our effective couplings do not coincide
exactly with their definitions. In particular, we differ in the
definition adopted for $g(q^2)$. Here, we have chosen to relate this
coupling directly to the $W^{\pm}$ propagator, in the same way that
the coupling $e(q^2)$ is related to the photon propagator.  They
choose, instead, the mixed propagator of the photon and the $W^3$
boson. Their convention follows from imposing that the tree level
relation among $e^2$, $g^2$ and the sine of the mixing angle,
$\sin^2\theta_W=s^2_W$, had to be maintained at the level of the
effective parameters, $i.e$,
\begin{equation}
\frac{s^2_W(q^2) g^2_{KL}(q^2)}{e^2(q^2)}=1\,, \label{senkl}
\end{equation}
where $s^2_W(q^2)$ is derived from neutral current processes.

In our case, to get a definition of $s^2_W(q^2)$ consistent with the
neutral current amplitude, we have to allow for the relation
(\ref{senkl}) to receive radiative corrections \cite{kuroda}, so that,
\begin{equation}
\frac{s^2_W(q^2) g^2(q^2)}{e^2(q^2)}=1+O(\hbar)\,.
\end{equation}
The coupling $g^2_{KL}(q^2)$ is related to a conserved current, and
hence it only receives contributions from loops of charged and
degenerate particles. There is, for example, no contributions due to
the Higgs. On the other hand, all the doublets under $SU(2)_L$
contribute to the coupling $g^2(q^2)$, and this includes the
Higgs. The coupling $g^2(q^2)$ is related with a $broken$ symmetry,
and this fact is not reflected in $g^2_{KL}(q^2)$. {\it We consider
that the definition of $g^2(q^2)$ is more appropriate for our
purposes, whereas $g^2_{KL}(q^2)$ is more appropriate to study the
neutral current and related processes}.

In the following sections, we will proceed to study the evolution with
scale of the effective couplings from \mz\ to the high energy region,
where we want to check for the validity of the unification
scenario. With this in mind, we see that the above definitions of the
effective couplings will be useful once we eliminate the bare
couplings in favor of the couplings given at the scale \mz.
Furthermore, bearing in mind that the combinations $\Pi^T_{ii}+2
\Gamma_i$ are proportional to $g_i^2$, we redefine $\Pi^T_{ii}+2
\Gamma_i= g_i^2 (\Pi^T_{ii}+2 \Gamma_i)$, and obtain: \widetext
\begin{equation}
\frac{1}{g_i^2(q^2)}=\frac{1}{g_i^2(m_Z^2)}+\left(\Pi^T_{ii}(q^2)+2
\Gamma_i(q^2) -\Pi^T_{ii}(m_Z^2)-2 \Gamma_i(m^2_Z)\right)\,.
\end{equation}
\narrowtext
\noindent As an example of unifying group we will take the minimal
choice, i.e., $G=SU(5)$.

\section{Threshold effects in the Standard Model.}
\label{sect3}
In this section we will study the Standard Model with minimal matter
contents, i.e., three generations of fermions and one scalar
doublet. From the point of view of unification, we do not expect that
the threshold effects included in the definition of the effective
couplings can change the negative results obtained with other
renormalization procedures \cite{amaldi}. This model is simply
intended to see how thresholds affect the evolution of gauge couplings
\cite{mia1}.

Since for scales $\mu \geq \mz$ we can regard all the fermions as
being massless, except for the top quark, the relevant masses for the
problem are \mz, \mw, \mt\ (top) and \mh\ (Higgs). The values of \mz\
and \mw\ are determined by experiments to be at \cite{data}:
$\mz=91.187 \pm 0.007\,GeV$, $\mw=80.2 \pm 0.3\,GeV$. But \mt\ and
\mh\ remain as free parameters, with lower experimental
bounds\footnote{ The recent data from the CDF Collaboration at
Fermilab \cite{top} indicate the existence of the top quark with
$\mt=174\pm 10^{+13}_{-12}\,GeV$. However, this data is still to be
confirmed, and therefore we would maintain the lower bound given in
Ref. \cite{data}. In any way, to fix or let the top mass free makes
only some very tiny differences in the numerical results, and it does
not influence the conclusions reached in this paper.}: $\mt \geq
131\,GeV$, $\mh \geq 60\,GeV$ \cite{mh}. On the other hand, imposing
as condition the validity of perturbation theory in the range of
scales considered, one gets the upper theoretical limit: $\mt,\;\mh
\leq 200 \,GeV$ \cite{maiani}. We will use in this section these
latter values for \mt\ and \mh.

The initial values of the couplings, $g_i^2(\mz)$, are identified with
the experimental values measured at the scale \mz:
\begin{eqnarray}
\alpha_3(\mz)&=&0.117 \pm 0.005 \,, \;\; \cite{data} \\
\alpha_e^{-1}(\mz)&=&127.9 \pm 0.3 \,\;\; \cite{ale} \\
\sin^2\theta_W(\mz)&=&0.2319 \pm 0.0005 \,\;\; \cite{data}.
\end{eqnarray}
In what follows, we will work with the ``constants" $\alpha_i=g_i^2/4
\pi$.  The initial value of \alp{2}\ is fixed by the relationship:
$\alp{2}(\mz)=\alp{e}(\mz) \sin^2\theta_W(\mz)$. And, as mentioned
above, the evolution of \alp{1}\ is calculated by the relation
$\alp{1}=3(\alp{e}-\alp{2})/5$, where the normalization factor is due
to the embedding of $U(1)_Y$ in $SU(5)$.

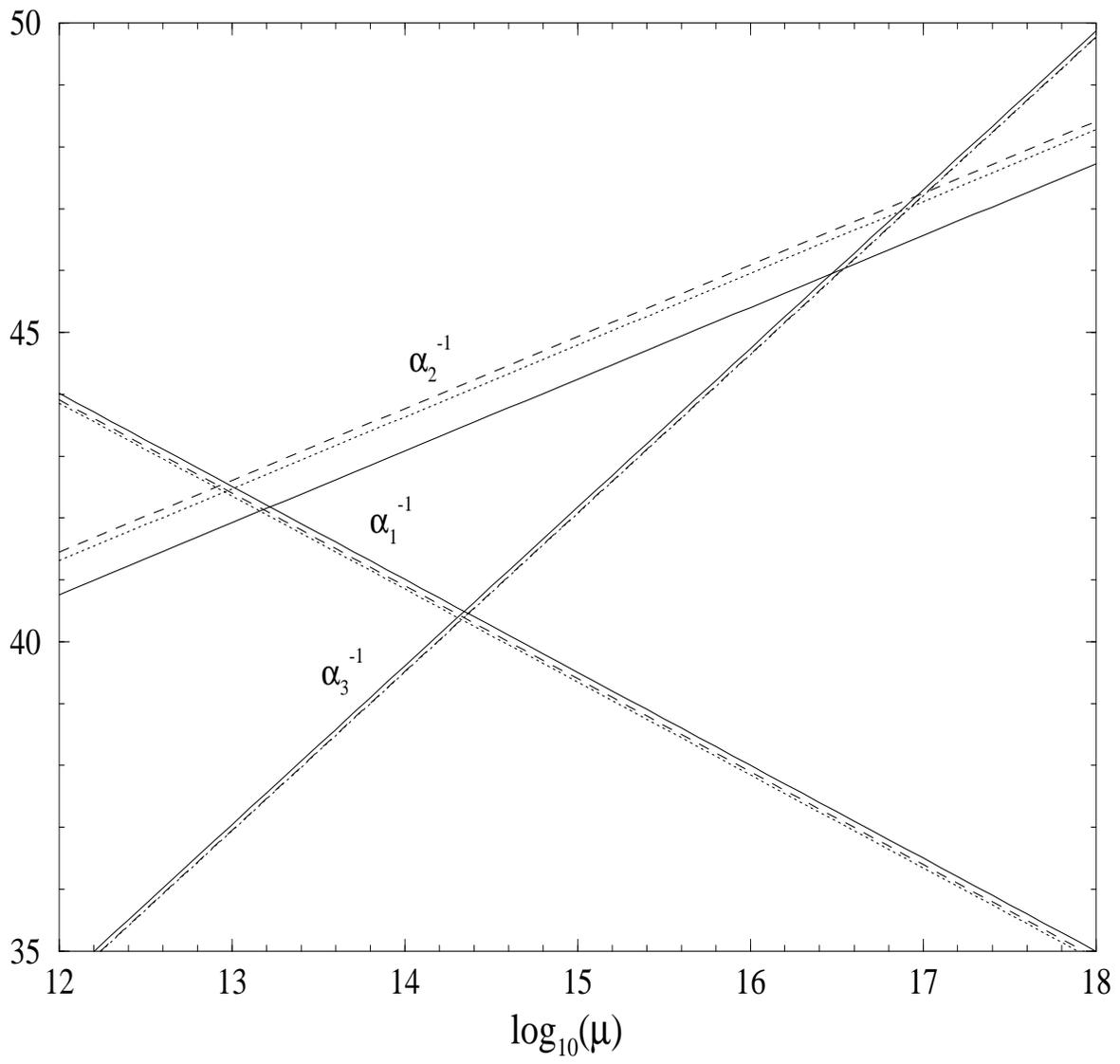
\begin{figure}\caption{ Evolution of the three couplings of the SM calculated
with the three 1--loop procedures: effective couplings (solid lines),
step--function (dashed lines), and $\overline{MS}$ (dotted lines). \label{sm1}}
\end{figure}

In Fig. (\ref{sm1}) we have plotted the effective couplings
(\al{i}{ef}) in the high energy region. We have also plotted the
couplings calculated in the step--function approximation
(\al{i}{\theta}), and using minimal subtraction (\al{i}{MS}). The
latter procedure is equivalent to the step--function approximation
when all masses are below or at the scale \mz. We see in the plot that
the approximation we use for \alp{1}\ and \alp{3}\ makes only very
little difference. Nevertheless, we can see that, due to threshold
effects produced by the top, \al{3}{ef} is a little larger than
\al{3}{\theta}, and this, in term, a little larger than \al{3}{MS}.
The top quark decouples at low energies ($\mt \ll 10^{12}\, GeV$), but
its effect propagates to the high energy region by the RGE, being the
decoupling in \al{3}{ef}\ smoother than in the
$\theta$--approximation.

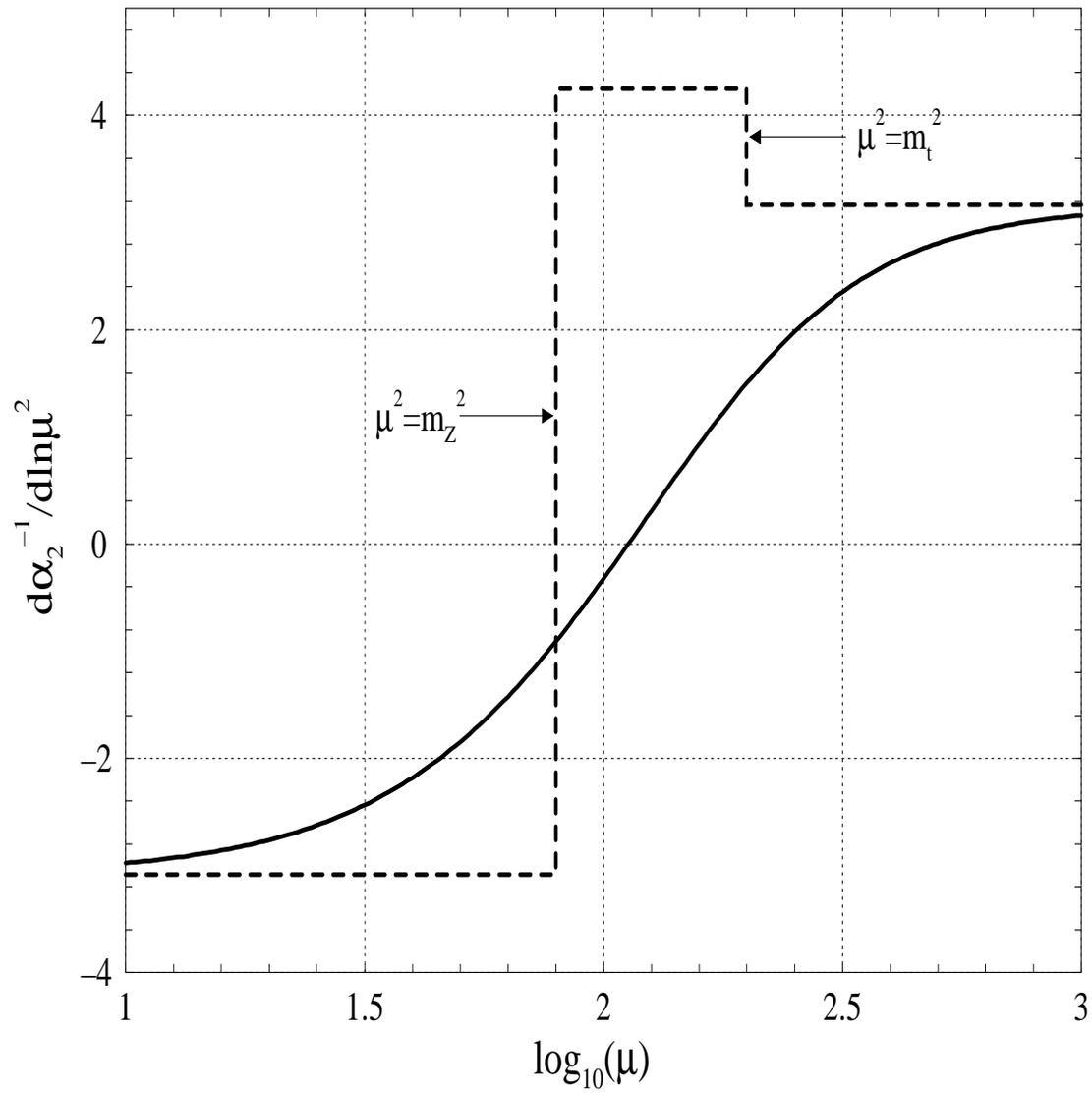
\begin{figure}\caption{ Derivative of the effective coupling
$\alpha_2^{-1}(\mu)$ respect  to
$\ln \mu$ (solid line). The dashed line shows the derivative within the
step--function approximation. \label{bet2}}
\end{figure}

 For the coupling \alp{2}, one would naively have expected that
the threshold contributions due to \mt\ and \mh\ contribute to
increase its slope. But these are not enough, because in this case
the dominant threshold effects are those due to the massive gauge
bosons, so that the slope of \al{2}{ef}\ decreases, as can be seen
from the plot. This effect is not present in \al{2}{\theta} (and
\al{2}{MS}) because we begin to integrate $precisely$ at the scale of
\mz. This can be seen clearly if we compare the derivatives, given
by:
\begin{eqnarray}
(4 \pi) \frac{\partial \alpha_2^{-1}\mid_{\theta}}{\partial \ln \mu^2}
&=& \frac{22}{3} \theta (\mu^2-\mzz) - \frac{1}{12}\theta
(\mu^2-m^2_h) -\theta (\mu^2-m^2_t)- \frac{37}{12} \,,\\
(4 \pi) \frac{\partial \alpha_2^{-1}\mid_{ef}}{\partial \ln \mu^2} &=&
\frac{13}{3}(s_{W}^2 f_g(0,a_{W})+c_{W}^2 f_g(a_{Z},a_{W})) +3
f_{\Gamma}(a_{W},a_{W}) \nonumber \\ & &-\frac{1}{12}
f_s(0,a_h)-f_f(0,a_f)-\frac{37}{12}\,,
\end{eqnarray}
\noindent where the functions $f_b(a_i,a_j)$ are defined in Appendix
A, and smooth out the ``step" of the Heaviside function, as demanded
by the Heisenberg uncertainty principle (see Fig. {\ref{bet2}). The
derivative of \al{2}{\theta} changes from negative to positive for
$q^2=\mzz$, while in the case of \al{2}{ef} threshold effects
postpone this change of sign \cite{ross}. Therefore, \al{2}{ef}
decreases when we begin the integration, and this makes the
differences with \al{2}{\theta} sizable only in the high energy
region.

The differences between step--function and effective couplings are in
that for the former, thresholds are considered as an ``instantaneous''
effect; we ``switch on'' and ``switch off'' them at a specific point
of the energy scale, i. e., at $q^2=m_i^2$. But indeed, thresholds are
spread over a certain range in the momentum scale. The threshold
crossing is not ``instantaneous''. As shown in Appendix A, only when
we begin to integrate at a scale $q_0^2 \gg m_i^2$ (massless
approximation), $or$ we end the integration at $q^2 \ll m_i^2$
(complete decoupling), we get the same contribution for the massive
degree of freedom $m_i$, in both \al{i}{\theta}\ and \al{i}{eff}. In
the general case, when $q_0^2 < m_i^2 < q^2$, even if we can apply the
limit $m_i^2 \ll q^2$, and then suppress all the terms $O(m_i/q)$ in
the threshold function contribution, we will get that,
$$
\al{i}{\theta} \sim \ln \frac{m_i^2}{q^2}\,,
$$
while,
$$
\al{i}{eff} \sim \sim \ln \frac{m_i^2}{q^2}- \ln C_i\,,
$$
where $C_i$ is a ``constant'' of $O(10)$\footnote{``Constant'' means
that this term can be approximated by a constant. See Appendix A for
details.}. This constant contribution is not more than a reminder of
the fact that, although if the particle is decoupled at $q_0$, in
going to $q$ we have to cross the region of scales $O(m_i)$, where the
degree of freedom is neither decoupled nor coupled\footnote{The word
``coupled'' is used here as opposite to ``decoupled'', i. e., maximum
contribution to the $\beta$ functions.}. Moreover, the transition
between the decoupled--coupled regimes, takes at least more than one
order of magnitude in the energy scale. Therefore, when beginning the
integration at the electroweak scale, \mz, it is not a good
approximation to consider the massive gauge bosons as completely
coupled.  }

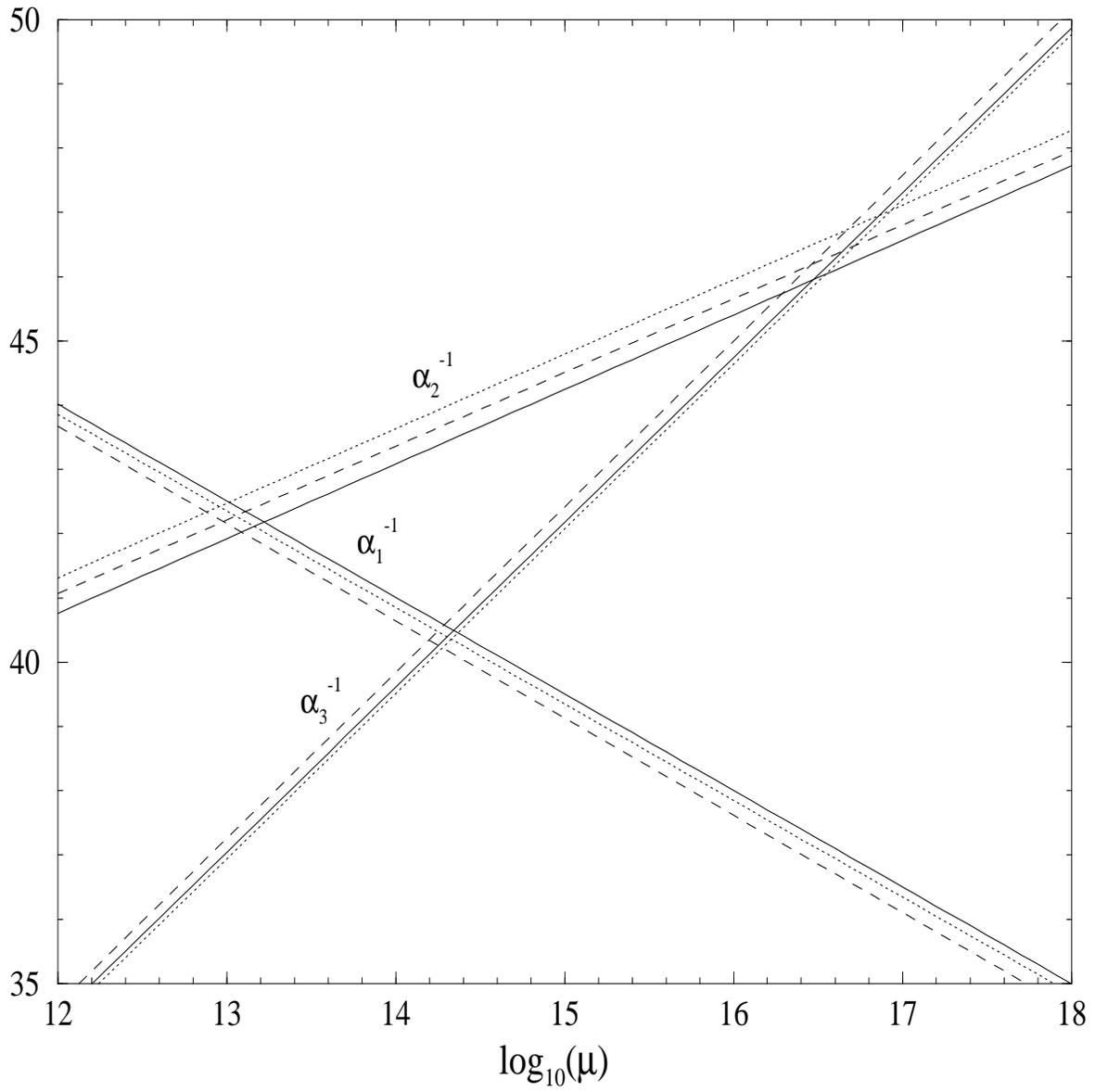
\begin{figure}\caption{ Evolution of the three couplings of the SM at 2--loop
order with $\overline{MS}$ (dashed lines), and 1--loop order with a mass
dependent method  (solid lines).
We also plot 1--loop $\overline{MS}$ for comparison (dotted
lines).\label{sm2}}
\end{figure}

We can now compare the corrections due to threshold effects with those
coming from improving the order of perturbation theory in the
mass--independent approach. In Fig. (\ref{sm2}) we have plotted
\al{i}{ef} at one--loop order, and \al{i}{MS} at two--loop order. We
also include \al{i}{MS} at one--loop order as a reference. We can now
see that two--loop effects are qualitatively similar to threshold
effects. Both of them raise \alp{3} and decrease \alp{2}, although for
\alp{2} the thresholds from the gauge bosons dominate over two--loop
effects. We can expect that, when we take into account both
corrections, calculating effective couplings at two loop order, the
differences relative to the values of \al{i}{MS} become more
accentuated.

Instead of doing the exact calculation, we have decided to study this
case in an approximate way.  We use it only as an indication of
which kind of behavior we can expect at 2--loop order with thresholds.
In Appendix B we describe in detail the argument followed to obtain
approximated two--loop effective couplings, based on the expressions
for the RGEs at 2--loops, and general properties of the threshold
functions.  When we study their evolution, we find what we expected:
\alp{2} decreases and \alp{3} increases slightly, with respect to the
one loop \al{i}{ef} values. Of course, these effects are not enough to
``unify" the three gauge couplings, but it is useful to keep them in
mind when studying theories with a larger presence of massive degrees
of freedom.

In spite of the no unification of the couplings, the Standard Model
with minimal matter contents has been useful to study the threshold
effects introduced in the effective couplings. In particular, we have
seen that qualitatively, the thresholds associated with the gauge
bosons are not negligible, contrary to what is assumed by the
step--function approximation (since we begin to integrate at the scale
$q^2=\mzz$). Quantitatively, moreover, this correction is larger than
the two--loop correction.

It is clear that thresholds are additive and they will become more
relevant the larger the number of degrees of freedom in the range
[\mz,$M_X$].  This is the case of the Minimal Supersymmetric extension
of the Standard Model (MSSM), that we will study in the following
section.

\section{Unification in the MSSM.}
\label{sect4}
 If supersymmetry is a symmetry of Nature, every particle of spin
``$j$" of the Standard Model must come with its supersymmetric (susy)
partner, of spin ``$j \pm 1/2$". Thus, we have gauginos, sleptons,
squarks and higgsinos in addition to the content of matter in the
Standard Model. We also need, at least, two Higgs doublets (with their
higgsinos) in order (a) to cancel the anomaly due to the higgsinos,
and (b) to give mass to the two components of the quark doublet. Since
the susy particles have not been detected, susy must be broken at low
energies, and therefore we have a larger number of arbitrary
parameters of masses to consider beyond the scale \mz.

One can try to constraint the susy spectrum in the context of
unification theories, making use not only of limits on proton decay
\cite{nath,hisano}, but also of cosmological arguments on the relic
abundance of the lightest susy particle (LSP) \cite{lsp}. In general,
the susy spectrum thus obtained lies below $1\,TeV$, and thus it is
therefore compatible with the stability of the hierarchy of scales in
the model under radiative corrections. In this sense, the limit of
``$1\,TeV$" is usually denoted as the ``naturalness" bound.

Most of the predictions have been obtained making use of the
step--function approximation to deal with susy thresholds. But the
more accurate description of threshold effects, such as given in the
effective couplings, could change the conclusions reached as we will
show in this section \cite{mia2}.

Of the susy spectrum, we will assume that susy is broken at the Planck
scale by a ``hidden" sector, so that the large number of susy masses
can be determined at the weak scale in terms of a small number of
parameters given at the unification scale \cite{lahanas}. Accordingly,
we take the following parametrization, neglecting the (higher order)
mixing between winos and higgsinos, and the s--tops left and right for
simplicity:

\noindent (a) For winos ($\tilde{w}$), and gluinos ($\tilde{g}$),
$$
\m{w}=m_{1/2}\;,\;\;\; \m{g}=3 m_{1/2}\,,
$$
\noindent (b) for squarks ($\tilde{q}$), and sleptons left
($\tilde{l}_L$) and right ($\tilde{l}_R$),
$$
m^2_i= m^2_{1/2}( c_i+\xi_0) \;, \;\;\; c_{\tilde{q}}=7\,,\;\;
c_{\tilde{l}}=0.5\,,\;\; c_{\tilde{r}}=0.15\,,
$$
where $\xi_0=(m_0/m_{1/2})^2$, $m_0$ being the common mass for the
scalars, and $m_{1/2}$ being the gaugino mass at the unification
scale; and

\noindent (c) for higgsinos, \m{h} will be taken as arbitrary.

Furthermore, there are five massive fields coming from the two Higgs
doublets: two charged Higgses ($m_+$), two neutral Higgses ($m_h$,
$m_H$), and one pseudoscalar ($m_a$), whose masses must satisfy the
relations: $m_+^2=m_a^2+m_W^2$, $m_H^2=m_a^2+m_Z^2-m_h^2$. The lighter
neutral Higgs is identified with the standard Higgs, and for the
remaining we take $m_+^2 \approx m_a^2 \approx m_H^2$. At the end, we
have as arbitrary mass parameters the following: \mt, \mh, $m_{1/2}$,
$m_0$, \m{h}\ and $m_H$. The value of $m_{1/2}$ is bounded from below
by the experimental searches for charginos \cite{data}, $m_{1/2}\geq
45\,GeV$. We take the remaining susy parameters at least of order \mz.

The expressions of the effective couplings given in Section
\ref{sect3}, with the functions $\Pi_{ii}^T+2 \Gamma_i$ calculated in
Appendix A, do not include the contribution of the susy degrees of
freedom. However, these are easily derived, taking into account that
(a) they do not appear in $\Gamma_i$, but only in $\Pi_{ii}^T$, and
(b) they are contributions from fermions or scalars (Eqs. (\ref{pif})
and (\ref{pis})).

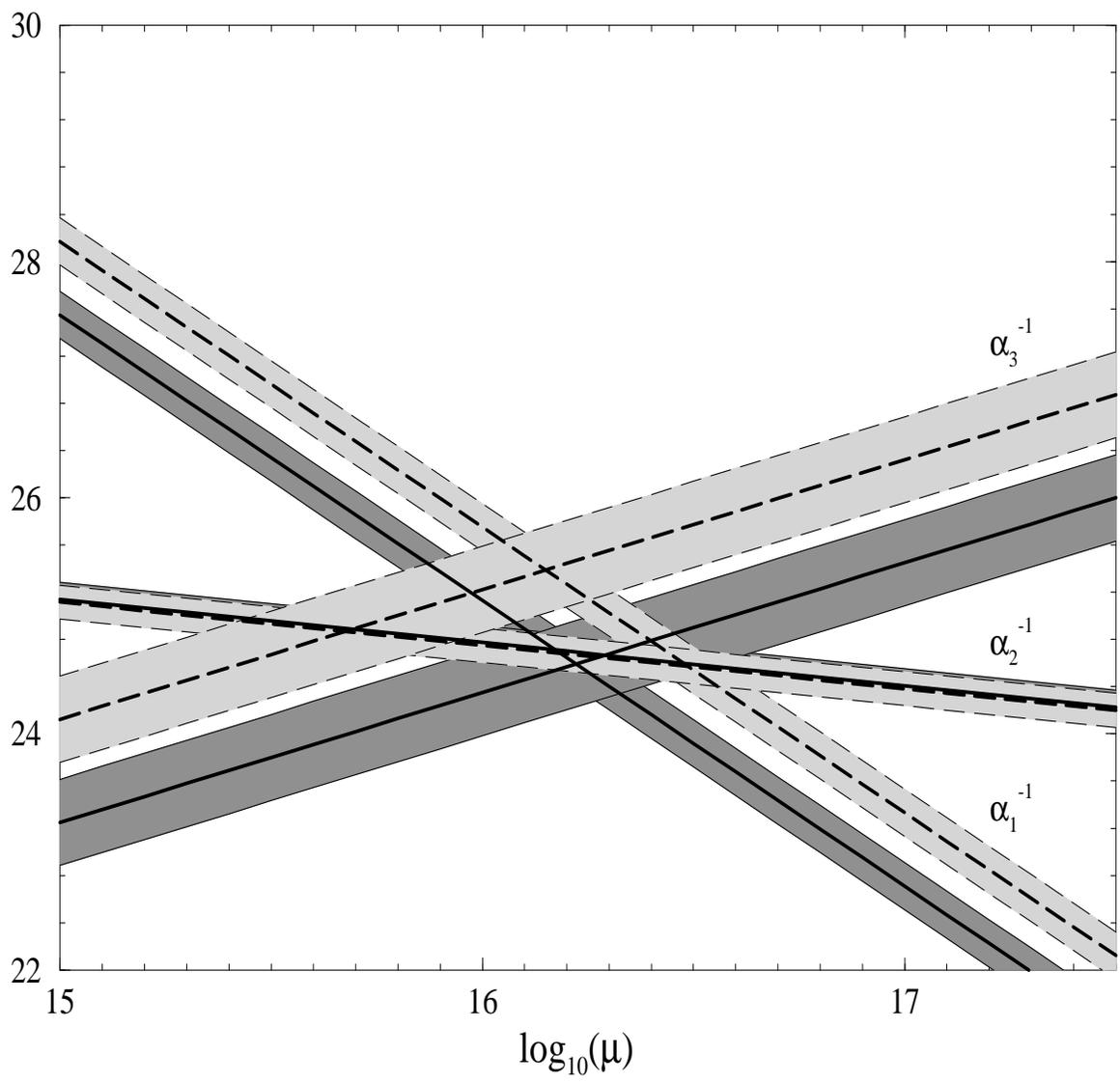
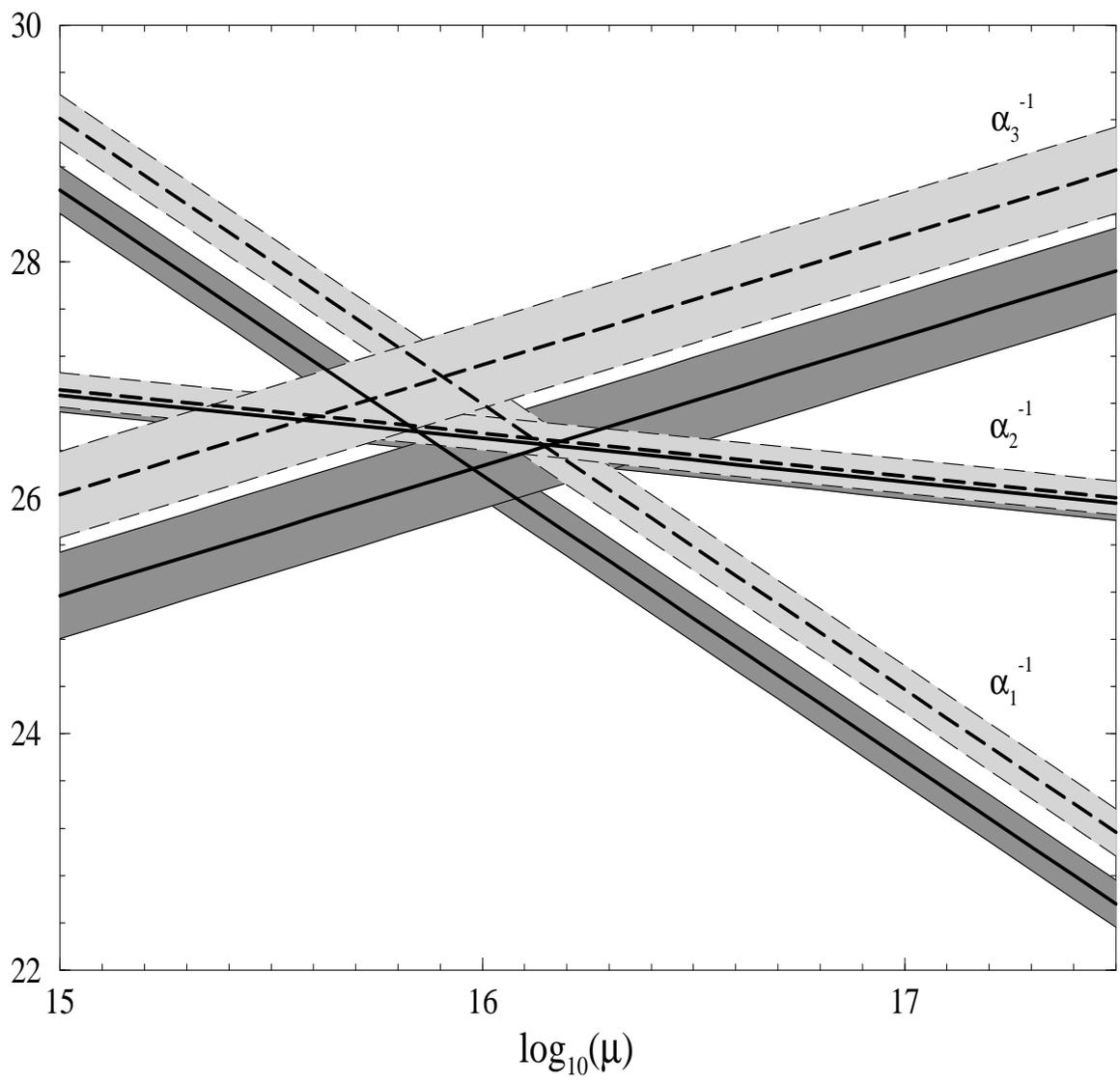
\begin{figure}\caption{ Evolution of the three couplings
of the MSSM at 1--loop order
calculated with  step--function (solid lines) and
 effective couplings (dashed lines), for different values of the susy mass
parameters:
 (a) $m_{1/2}= 45\; GeV$ and $m_0=m_Z$,
 (b) $m_{1/2}=m_0=1\;TeV$;
we take $m_{\tilde{h}}=m_{+}=m_H=m_0$ and $m_t=m_h=200 \;
GeV$. We include in the plot the experimental error band for each coupling.
\label{susy1}}
\end{figure}

As a first attempt to look for the unification of the effective
couplings, we simply study their evolution with scale, without making
any reference to the unifying group.  In Fig. (\ref{susy1}.a) and
(\ref{susy1}.b) we have represented \al{i}{ef} for two different
values of the susy parameters (the lower and upper bound). In order to
make a meaningful comparison, we have also included \al{i}{\theta}
calculated with the same susy parameters.  Taking into account the
experimental errors in the coupling constants, we see that the
\al{i}{\theta} unify (they $cut$ at one point, $M_X$, of the energy
scale), for both values of the susy parameters. Due mainly to the
experimental error in $\alpha_3 (\mz)$, at this level of approximation
one can not extract much more information about the susy spectrum with
the \al{i}{\theta}.

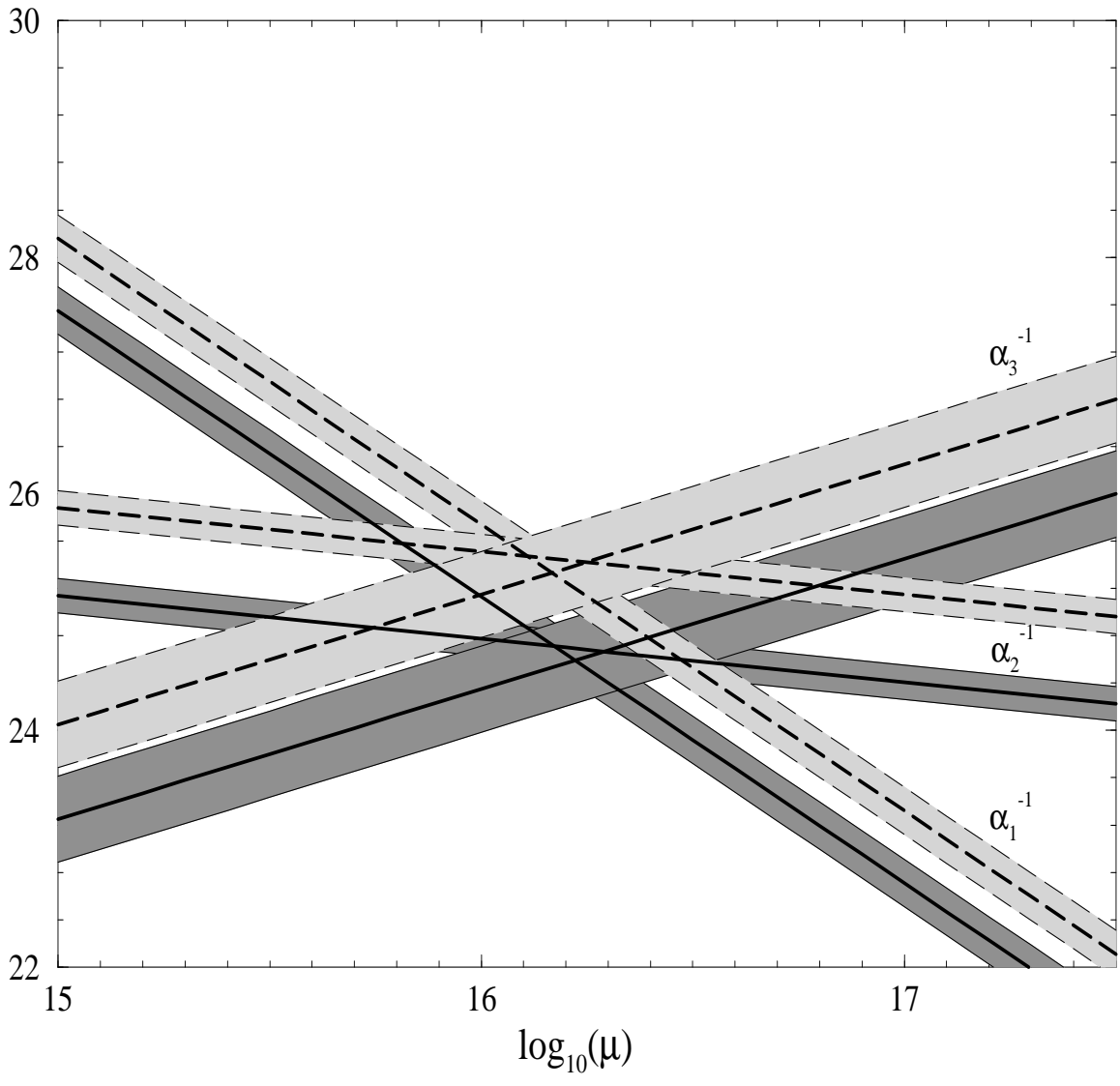
\begin{figure}
\caption{ Same as Fig. 5.a, but now without including the threshold
effects due to the massive gauge bosons in the evolution
of $\alpha_2^{-1}$.
\label{susytp1}}
\end{figure}

But the situation is not as unambiguous for the \al{i}{ef}: to get
unification, effective couplings $prefer$ higher values for the susy
parameters, around the naturalness bound of $1\,TeV$.  Now, the
threshold effects due to the gauge bosons play an important role in
reaching or not unification. While the contribution of susy masses
(fermions and scalars) tends to increase the slope of \al{3}{ef} and
\al{1}{ef}, for the coupling \al{2}{ef} this effect is compensated by
the decrease of the slope due to the gauge thresholds. Because of
this, the differences between \al{2}{ef} and \al{2}{\theta} are almost
negligible, contrary to those obtained for the other two couplings,
making \al{2}{ef} move away from the ``crossing'' point with the other
two effective couplings. Indeed, if the bosons $W^{\pm}$ and $Z^0$
were massless, the slope of \al{2}{ef} would increase sufficiently so
as to allow the unification of the three couplings even for a susy
spectrum of $O(\mz)$ (see Fig. (\ref{susytp1})).  $Qualitatively$,
we will achieve the same conclusions as we would if using the
step--function approximation: the perturbative unification of the
couplings is compatible with a susy spectrum below $1\,TeV$, and a
unification scale $M_X$ high enough to fulfill the experimental lower
limit on proton decay.  However, when we take into account the
thresholds, including those of $W^{\pm}$ and $Z^0$, this general
conclusion is slightly modified, and we get more information about the
susy spectrum: it can not be of $O(\mz)$.  We see that the
electroweak breaking, being a ``low energy'' process (order \mz), has
effects on processes which take place at ``high energy'' (order of
$M_X$), such as the perturbative unification of the couplings.

 Simply plotting the evolution of the couplings is not the best
way for making predictions, and even less so with the big uncertainty
we have in the experimental determination of $\alpha_3(\mz)$.
Indeed, of the three gauge couplings the value of \alz\ is the worst
known experimentally, in the sense that there is not a general
agreement between different measurements
\cite{dis,lepqcd1,lepqcd2}. In general, the values obtained from
experiments at low energies (below \mz) are lower than the values
inferred from the decay of the $Z^0$ into hadronic states (LEP data)
(see Table \ref{table3}). Because of this discrepancy it is better not
to take any value of \alz\ as the initial data and, instead, {\it to
derive} it from the unification condition:
\begin{equation}
\alp{1}(M_X)=\alp{2}(M_X)=\alp{3}(M_X)=\alp{G}\,.
\label{unif1}
\end{equation}
In this way, susy masses can be bounded by demanding that $0.108 \leq
\alz \leq 0.125$, which are the experimental lower and upper values
quoted in Table \ref{table3}.

  Solving the system of equations derived from (\ref{unif1}), one
gets the general expressions:
\widetext
\begin{eqnarray}
\alzi \mid_{\theta} &=& \frac{1}{7} (15 \alp{2}(\mz)-3 \alp{e}(\mz)) +
                    \frac{1}{56 \pi} \sum_i d_i f( \frac{m_i}{\mz})
                    \,,
\label{al3} \\
\ln \frac{M_X}{\mz} \mid_{\theta} &=& \frac{\pi}{14}(3 \alp{e}(\mz)-8
\alp{2}(\mz)) -\frac{1}{84} \sum_i d_i' f( \frac{m_i}{\mz}) \,,
\label{mx}
\end{eqnarray}
\narrowtext
\noindent where the sums run over $all$ the masses (susy and
non--susy).  When we work within the step--function approximation, the
functions ``$f$'' are simply given by $\ln (m_i^2/ m_Z^2)$; thus ,
only masses $m_i$ greater than \mz\ contribute. Within the effective
couplings approach, and making use of the approximation for the
threshold function given in Appendix A (Eq. (\ref{aaprox})), we get
\begin{equation}
f (m_i/\mz)=\ln \frac{\mzz +c_i m_i^2}{\mzz} \,,
\label{appt}
\end{equation}
where the value of the $c_i$ mostly depends on the kind of massive
particle running in the loop (see Table \ref{consc}). As we have seen
in Sect. 3, and also in Appendix A, this constant is intended to
smooth the threshold--crossing, and to control, not only beyond but
also $below$, how far it is the threshold from the electroweak scale.

Therefore, we see that, independently of whether we use effective
couplings or the step--function approximation, the behavior with susy
masses of \alzi\ and $\ln M_X$ will be the same: the higher the susy
masses, the higher \alzi\ and the lower is $\ln M_X$. But now, with
the effective couplings we have also a non zero contribution of the
massive gauge bosons, which are explicitly given by:
\widetext
\begin{eqnarray}
\alpha_3^{-1}(\mz) &\rightarrow& \frac{1}{28 \pi}\left( 39
s^2_W\ln\frac {\mzz+2 \mw^2}{\mzz+5 m^2_W}- 52\ln\frac {\mzz+5
\mw^2}{\mzz}-36\ln\frac {\mzz+2.5\mw^2}{\mzz}\right) \approx -1.44
\,,\\
\ln M_X &\rightarrow& \frac{1}{168}\left( 26 s^2_W\ln\frac {\mzz+2
\mw^2}{\mzz+5 m^2_W}+ 65\ln\frac {\mzz+5 \mw^2}{\mzz}+45\ln\frac
{\mzz+2.5\mw^2}{\mzz}\right) \approx +0.88 \,.
\end{eqnarray}
\narrowtext Thus, these terms decrease the value of \alzi\ and
increase the value of $\ln M_X$, with respect to those obtained in the
step--function approximation.

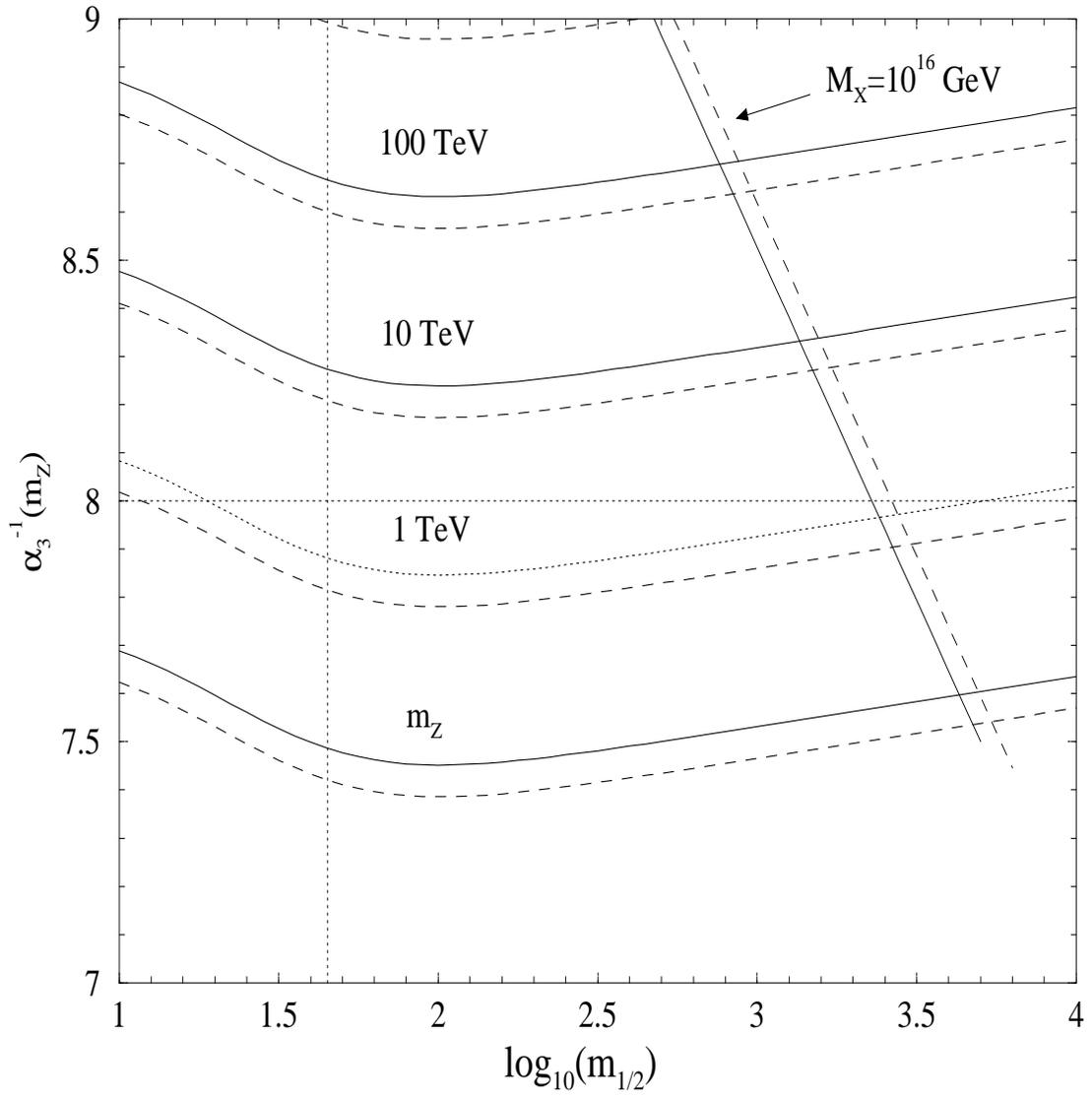
\begin{figure}\caption{ Values of $\alpha_3^{-1}(m_Z)$,
compatible with the unification condition (Eq. 19), calculated with
effective couplings at 1--loop order, for different values of
$m_{\tilde{h}}=m_H$: $m_Z$, $1\;TeV$,  $10\,TeV$,  $100\;TeV$. Dotted lines are
the experimental limits on $\alpha_3^{-1}(m_Z)$ and $m_{1/2}$; solid lines are
for $m_t=m_h=200\;GeV$, and dashed lines for $m_t=91\; GeV$ and $m_h=60\; GeV$.
The straight lines (solid for $m_t=m_h=200\;GeV$ and dashed for $m_t=91\; GeV$,
$m_h=60\; GeV$)  are the upper limit obtained for
$\alpha_3^{-1}(m_Z)$ when imposing $M_X=10^{16}\;GeV$ ($m_{\tilde{h}}=m_H$
increase along these lines from bottom to top). The allowed region for
$\alpha_3^{-1}(m_Z)$ are to the left of the straight lines, and between the
dotted lines,  $8 \leq \alpha_3^{-1}(m_Z) \leq 9.2$ and
$m_{1/2} \geq 45\, GeV$.
\label{susyt3}}
\end{figure}

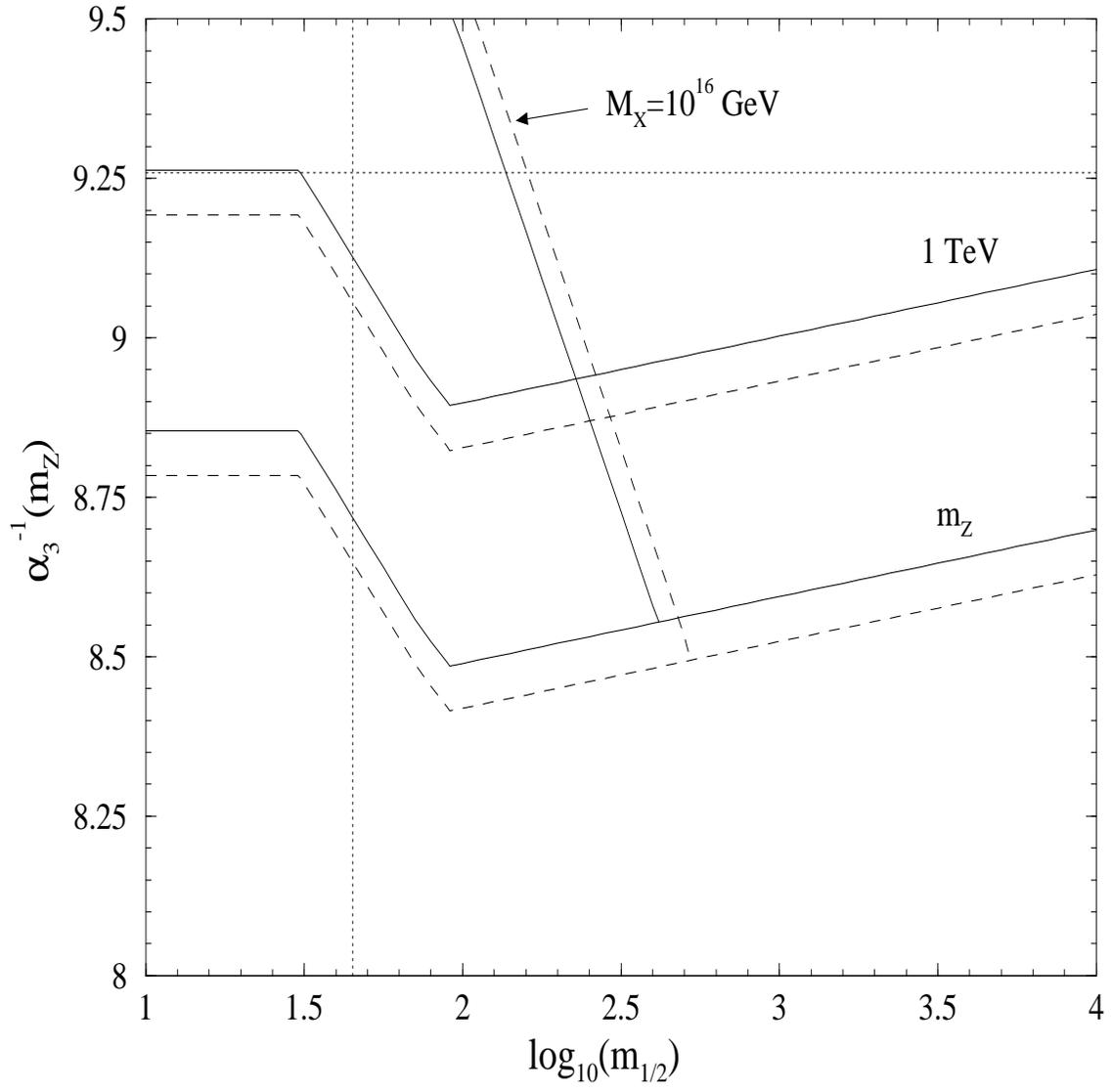
\begin{figure}\caption{ Same as Fig. 7, but with $\alpha_3^{-1}(m_Z)$
calculated
with the step--function approximation. \label{susyp3}}
\end{figure}

This behavior is clearly seen when we compare Fig. (\ref{susyt3})
(effective couplings) and Fig. (\ref{susyp3}) (step--function), where
we have plotted \alzi\ versus $log_{10}(m_{1/2})$, for different
values of the remaining mass parameters. The allowed region in the
map $\alzi-log_{10}m_{1/2}$ is delimited by the experimental bounds on
\alzi\ and $m_{1/2}$ given above, and also by the theoretical limit
$M_X \geq 10^{16}\,GeV$, which is a safe bound for proton decay. Since
higgsinos and heavy Higgses contribute with the same sign, we have
simply taken $\m{h}=m_H$.  Moreover, \alzi\ and $\ln M_X$ depend only
very slightly on the parameter $\xi_0=(m_0/m_{1/2})^2$, so that we
have taken $\xi_0=1$ as a typical value for the plot.  As we
increase the mass of the higgsinos, we decrease (increase) the value
of \alz\ (\alzi) compatible with unification. However, since the
step--function procedure does not distinguish masses lower than \mz\,
with $any$ higgsino mass below \mz\, and gauginos masses of $O(\mz)$,
we get the limit $\alz < 0.121$, perfectly compatible with the
experimental band for this coupling.  As we have previously seen, no
constraint on the susy spectrum is obtained at this order.  On the
other hand, with the effective couplings we would get the upper value
of \alz\ in the limit $\m{h} \ll \mz$, but, now, due to the
contribution of the gauge bosons, this limit has no physical interest
($\alz < 0.140$). The point is that also for higgsinos masses of
$O(\mz)$, and almost for a susy spectrum of $O(1\, TeV)$, the value we
get for \alz\ is too high. Turning the argument upside down, if we
demand $\alz \leq 0.125$ and $M_X \geq 10^{16}\, GeV$, we get a lower
bound on $\m{h}=m_H$, and an upper bound on $m_{1/2}$. These bounds
depend on \mt\ and \mh\ (the lower \mt\ and \mh, the higher $m_{1/2}$
and $\m{h}$), and also on $\xi_0$, although these bounds remain
practically unchanged for $\xi_0 \geq 10^4$.  For this value of
$\xi_0$ we get:
\begin{eqnarray}
m_{1/2}&\leq& 7\, TeV \;\;\;\;
(m_h=60\,GeV\,,\;\;m_t=91\,GeV) \;,\nonumber \\ \m{h}&\geq& 370 \,
GeV\;\;\; (m_h=m_t=200\,GeV\,,\;\;\m{h}=m_H) \;.
\end{eqnarray}
By giving these bounds and the above bounds on \alz, we have taken
into account the experimental errors in $\sin \theta_W^2 (\mz)$ and
$\alpha_e^{-1}(\mz)$. For the central values of these quantities we
would obtain: $ m_{1/2} \leq 3.8 \, TeV$, $\m{h} \geq 1\, TeV$. Any
kind of prediction from unification is very sensitive to small
variations in the input parameters, and even small errors in the
experimental data induce serious uncertainties\footnote{ Note that the
values quoted in \cite{mia2} are different, due to the different value
of $\sin^2 \theta_W(\mz)$ considered in that work.}.

With this calculation we have recovered in a more detailed way the
general results we obtained before. At the one--loop order, we can get
the perturbative unification of the couplings, compatible with the
naturalness bound for the susy masses and proton decay--lifetime
limits, independently of whether we take or not threshold effects into
consideration. But the inclusion of more accurate thresholds than
those described by the step--function, drives the susy spectrum close
to the $1\, TeV$ bound and the value of \alz\ into the range of LEP
data.

{\it Threshold effects as considered here are not higher order
corrections, but a correction to the simplest step--function
approximation}. As a first approach, just at one--loop order, this
latter procedure gives a good indication for unification with a viable
susy spectrum. Nevertheless, when one improves the treatment of
thresholds, working with effective couplings, although the conclusions
are not drastically changed, the result at one--loop does not leave
much room for more corrections, for example 2--loop order corrections.

  In fact, it is not necessary to carry out a complete calculation
of thresholds at two--loop to get the general behavior. It is enough
to take into account that two loop effects tend to lower the value of
\alzi\ by about 10 \%, for both the step-- function and effective
couplings\footnote{ For the effective couplings, this can be easily
shown by resorting to the kind of approximation given in Appendix B.}.
  In the first case, this effect drives the upper limit obtained at
one--loop order towards the experimental upper limit ($\alz \leq
0.125$), while the naturalness bound on susy masses gives now the
lower bound $\alz \geq 0.121$. With the effective couplings, a 10\%
decrease pushes the values of \alzi\ obtained with $\m{h}=m_H \leq
1\,TeV$ {\it away from the experimental band}.  We would need masses
of order $10\,TeV$ to get $\alz \leq 0.125$. For the two--loop
calculation, the dependence on the parameter $\xi_0$ is not so mild,
and the value of \alzi\ increases with it. In any case, to have
$\m{h}=m_H$ in the range of $TeV$, we would need $\xi_0 \geq 10^{10}$,
i.e., squarks and sleptons of order $1\,PeV$.  Therefore, we can not
conclude, at two--loop order, that the simple scenario for
perturbative unification is $viable$; in the sense that it is
not $compatible$ with the experimental data on \alz\ and the
naturalness bound on the susy spectrum.

 The question still remains of the corrections due to the heavy
degrees of freedom associated to the grand unification group (in our
case $G=SU(5)$). As we have seen previously, light thresholds
introduce non negligible differences in the evolution of the running
couplings, and the same will take place with the heavy masses ($M_j$),
mainly when the ratio $\mu/M_j$ begin to approach to one. The
evolution of the effective couplings, including only the light degrees
of freedom, is valid up to scales $\mu \ll M_j$, where we are sure
about the decoupling of the heavy masses $M_j$. The relationship
between the couplings $\alp{i}(\mu)$ and $\alp{G}(\mu)$ is obtained
integrating out the heavy degrees of freedom from the complete action
$S[G]$, and is given by:
\begin{equation}
\alp{i}(\mu)=\alp{G}(\mu)+\lambda_i(\mu,M_j) \,. \label{unif2}
\end{equation}
The functions $\lambda_i(\mu,M_j)$ include logarithmic contributions
of the heavy masses, as well as a constant term due to the light
degrees \cite{lam2}.   As we point out in Sect. 3, the
logarithmic contribution is what we get when aplying the limit $M_j
\gg \mu$ to the threshold functions.

For non--susy $SU(5)$, we have three basic parameters to describe the
heavy masses: $M_V$ (gauge bosons), \mfi\ (colored Higgs) and \msi\
(heavy scalar in the adjoint representation). For susy $SU(5)$ we have
to add the associated susy partners. If we assume, as it is common,
that the members of each heavy supermultiplet are degenerated, we
get\footnote {The expressions given in Ref. \cite{mia2} for
$\lambda_3(\mu,M_j)$ and also $\ln M_V$ included minor mistakes that
we have corrected here (see Eq.  (\ref{lam3}) and (\ref{mv})). We have
also corrected the constant term due to the light degrees of freedom
that appears in $\lambda_i$. In Ref. \cite{mia2} we reproduced the
terms given by N.-P. Chang et. al. in \cite{lam2}. Here, we have
recalculated them to be compatible with our definition of effective
couplings. Anyway, the numerical differences are small, and they do
not affect the conclusions.}:
\begin{eqnarray}
(4
\pi)\lambda_1(\mu,M_j)&=&\frac{66}{5}+\frac{96}{5}\ln\frac{M_{V}}{\mu}-
\frac{4}{5}\ln\frac{M_{\Phi}}{M_{V}}-
\frac{20}{3}\ln\frac{M_{\Sigma}}{M_{V}} \,,\\
(4 \pi)\lambda_2(\mu,M_j)&=&\frac{41}{6}+8\ln\frac{M_{V}}{\mu}-
	                   8\ln\frac{M_{\Sigma}}{M_{V}} \,,\\
(4 \pi)\lambda_3(\mu,M_j)&=&\frac{5}{4} -2\ln\frac{M_{\Phi}}{M_{V}}
                   -\frac{26}{3}\ln\frac{M_{\Sigma}}{M_{V}}
                   \label{lam3}\,.
\end{eqnarray}
These expressions are valid at one--loop order for scales $\mu$ such
that $m_i \ll \mu \ll M_j$. Since heavy masses are typically of order
$10^{16}\,GeV$, and light masses are expected to be less than
$1\,TeV$, we choose $\mu=10^{7}\,GeV$.

As before, from (\ref{unif2}) we get a system of equations, where now
the unknowns will be \alp{G}, \alzi, and $\ln M_V$ instead of $\ln
M_X$. We get for \alzi\ and $\ln M_V$ the following expressions:
\begin{eqnarray}
\alpha_3^{-1}(\mz)&=& \frac{1}{2}\left(3\alpha_2^{-1}(\mz)-
  \alpha_1^{-1}(\mz)\right)+\frac{1}{8\pi}\sum_i g_i F_i(m_i,\mu)
  \nonumber \\ & & -\frac{3}{5\pi}\ln
  \frac{M_{\Phi}}{\mu}-\frac{3}{5\pi}\,,
\label{alz3}\\
\ln M_{V} &=&\frac{3\pi}{8}\left(\alpha_1^{-1}(\mz)-
  \alpha_2^{-1}(\mz)\right)+\frac{3}{32}\sum_i g'_i F_i(m_i,\mu)
  \nonumber \\ & & +\frac{3}{40}\ln \frac{M_{\Phi}}{\mu}
  -\frac{1}{8}\ln \frac{M_{\Sigma}}{\mu} -\frac{191}{320}\,,
\label{mv}
\end{eqnarray}
where, as before, the sums run over all the light masses.  These
expressions are valid for both effective couplings and step--function
approximation. In the first case, the $F_i(m_i,\mu)$ are the
approximated threshold functions defined in (\ref{approx}); in the
second case, we simply take $F_i(m_i,\mu)=\ln (m_i^2/\mu^2)$ when $m_i
\geq \mz$, and $F_i(m_i,\mu)=\ln (\mzz/\mu^2)$ when $m_i <
\mz$. Notice that for \al{3}{\theta} and $\ln M_V\mid_{\theta}$, the
dependence on the scale $\mu$ is exactly canceled out at one--loop
order. For the effective couplings, this dependence remains negligible
if we maintain the condition $m_i \ll \mu \ll M_j$.

The qualitative behavior of \alzi\ and $\ln M_V$ with susy masses is
the same observed without taking into account the heavy thresholds.

Our interest now shifts to the heavy mass parameters \mfi\ and \msi.

Since the value of \alzi\ only depends on \mfi, the limits on this
parameter will put bounds on \alzi. The lower bound on \mfi\ comes
from the experimental limits on proton decay via dimension five
operators. The lifetime for the dominant mode is given by
\cite{hisano}: \widetext
\begin{equation}
\tau(p\rightarrow K^+\bar{\nu}_{\mu})=6.9\times10^{31}\left|
 \frac{0.003}{\beta}\frac{\sin2\beta_H}{1+y^{tK}}\frac{M_{\Phi}}{10^{17}}
 \frac{10^{-3}}{f(\m{q},\m{q},\m{w})+f(\m{q},\m{l},\m{w})}\right|^2
 yrs\,,
\end{equation}
\narrowtext
\noindent where we have introduced three more unknown parameters: the
hadron matrix element parameter $\beta$, which ranges from 0.003 to
0.03 $GeV^3$; the ratio of the vacuum expectation values of two Higgs
doublets, $\tan \beta_H$; and the parameter $y^{tK}$, which represents
the ratio of the contribution of the third generation relative to the
second generation to proton decay. To allow an \mfi\ as low as
possible, we take $\beta=0.003\,GeV^3$, $\sin 2\beta_H=1$ and $\mid
1+y^{tK}\mid=1$. The experimental limit for this mode is $\tau (p
\rightarrow K^+ \bar{\nu}_{\mu}) >1.0 \times 10^{32}\, yr$
\cite{data}, and this translates into the following lower bound on
\mfi: \widetext
\begin{equation}
\mfi> 1.2\times 10^{20} \left(
f(\m{q},\m{q},\m{w})+f(\m{q},\m{l},\m{w}) \right) = \mfi^{min}\,.
\label{phimin}
\end{equation}
\narrowtext
\noindent The functions $f(m_1, m_2, \m{w})$ come from the dressing of
the dimension--five operator with the wino exchange, needed to convert
it into suitable four--fermion operators for proton decay
\cite{pd5}. With the parametrization we have adopted for susy masses,
the combination that appear in (\ref{phimin}) is given by: \widetext
\begin{eqnarray}
\lefteqn{f(\m{q},\m{q},\m{w})+f(\m{q},\m{l},\m{w})=} \nonumber \\ &
&\frac{1}{6.5m_{1/2}} \left(\frac{\xi_0+13.5}{\xi_0+6}\ln (\xi_0+7)-
\frac{\xi_0+0.5}{\xi_0-0.5}\ln (\xi_0+0.5)\right)\,.
\end{eqnarray}
\narrowtext
\noindent So that, $\mfi^{min}$ depends on \msw\ and \sio, decreasing
with both of them.

On the other hand, the upper limit on \mfi\ and \msi\ is derived by
requiring that the Yukawa couplings involving these fields remain as
perturbative couplings below the Planck scale. This leads to the
conditions $\mfi \leq 2M_V$ and $\msi \leq 1.8 M_V$ \cite{hisano},
which combined with (\ref{mv}) give: \widetext
\begin{eqnarray}
\ln M_{\Phi}&<&
  \frac{15\pi}{37}\left(\alpha_1^{-1}(\mz)-\alpha_2^{-1}(\mz)\right)
  +\frac{15}{148}\sum_i g`_i(m_i,\mu) \nonumber \\ & &-\frac{5}{37}\ln
  M_{\Sigma} -\frac{191}{80}+\frac{40}{37}\ln 2=\ln M_{\Phi}^{max} \,.
\label{phimax}
\end{eqnarray}
\narrowtext
\noindent The upper limit on \mfi\ depends not only on the susy masses
but also on \msi.

We have already all the ingredients (Eq. (\ref{alz3}), (\ref{phimin})
and (\ref{phimax})) to check if the scenario of perturbative
unification is compatible with all the constraints on \alzi, \mfi\ and
the susy masses.  In order to compare, we examine first the results
obtained with the step--function approximation. In this case, there is
no problem in having \alz\ within its experimental range and the susy
masses below $1\,TeV$, as it can be seen in Table \ref{susypt}.

Some comments about these data. The bounds on the heavy and light mass
parameters are derived from imposing the condition $\mfi^{min}(\msw,
\sio) \leq \mfi^{max}(m_i,\msi)$. In this way, once we calculated the
lower bound on \mfi\ from (\ref{phimin}) with the maximum allowed
value for \sio\ (determined by $m_0^{max}$ and $\msw=45\,GeV$), this
lower bound on \mfi\ gives us the maximum allowed value for \msi. At
the same time, the minimum value of \msi\ gives us, on the one hand,
the lower bound on \sio\ (upper values on \msw\ and $m_0$), and also
gives us the upper bound on \mfi\ (Eq.  (\ref{phimax})) and the lower
bound on $m_0$. These latter bounds are derived taking $\msw=45\,GeV$
and \m{h}=\mz, and therefore are independent of any other constraints
on the susy masses. The same does not occur with the other bounds,
which depend on the value of the upper limit for the susy masses being
considered (mainly $m_0^{max}$).  (We have also taken $\mt=91\,GeV$
and $\mh=60\,GeV$ to get the values of the Table, except to derive
$\alz^{(min)}$).

In principle, we do not have any constraint on $\msi^{min}$, except
 the requirement that there is no large splitting between heavy
 masses. As \mfi\ and $M_V$ will be around $10^{16}\,GeV$, we have
 taken $\msi^{min}=10^{13}\,GeV$ to give the results quoted in Table
 \ref{susypt}.  With this choice, we see that when imposing the
 naturalness bound on the susy masses, we get $\msw < \mz$, near its
 lower experimental limit, and also narrow ranges for $m_0$, \mfi\ and
 \msi: $m_0\approx 1\,TeV$, $\mfi \approx 10^{16.7}\,GeV$ and $\msi
 \approx 10^{13}\,GeV$. We can increase the ranges of \msw\ and \m{h}
 if we take a lower value of \msi. For example, with
 $\msi^{min}=10^{10}\,GeV$ we obtain: $45\,GeV \leq \msw \leq 610 \,
 GeV$, $415 \,GeV \leq m_0 \leq 1\,TeV$. In any case, if we maintain
 the naturalness bound of $1\,TeV$ for the susy masses, it is $not$
 possible to maintain all heavy masses of the same order. Since the
 value of \mfi\ is practically fixed by the experimental limits on
 proton decay, and $M_V$ diminishes with \msi, the most favorable
 situation will be: $\msi < M_V \alt \mfi$.

When we calculate with effective couplings, the values of \alzi\ are
lower than those obtained with the step--function approximation,
mainly due to the thresholds of the massive gauge bosons. An now, to
satisfy the constraints on \alz\ and \mfi\ we need to have $m_0$,
$\m{h}$ or $m_H$ beyond $1\,TeV$. The results derived in this case are
presented in Table \ref{susytt}. We see that the lower $\xi_0$ (i.e.,
the lower squark and slepton masses), the higher will be \mfi, and
therefore higher values of \m{h}\ or $m_H$ will be needed to get $\alz
\leq 0.125$. If squarks and sleptons are below $1\,TeV$, the Higgs and
higgsino masses have to be beyond this bound. Notice that this
conclusion is not affected by the value adopted for $\msi^{min}$. It
is the value of $\mfi^{min}$ and the condition $\alz=0.125$ which
fixes the lower bound for $\m{h}=m_H$, independently of \msi.

As before, the bounds on the susy masses depend very midly on the
values of \mt\ and \mh, decreasing as \mt\ and \mh\ increase. Thus, we
have taken the values $\mt=\mh=200\,GeV$ to give the results in Table
\ref{susytt}.

We have seen that, in the context of perturbative gauge coupling
unification at one loop order, the use of effective couplings and the
requirement of having susy masses not too high, $favor$ a value of
\alz\ near its upper experimental bound, and this is independent of
whether or not we include the heavy threshold effects.  However,
when we try to incorporate corrections at two--loop order, or when we
take into account heavy threshold effects, the maximum allowed value
for \alz\ is not enough to get unification, and we have to increase
the susy masses beyond their naturalness bound.  Therefore, when we
use effective couplings is not so easy to maintain the naturalness
bound for the susy masses, as it was when we treat the thresholds with
the step--function approximation.

\section{Comparison with other approaches and conclusions.}
 In this paper we have studied how threshold effects associated to
massive degrees of freedom modify the evolution of gauge coupling
constants with scale; we have done it for both the Standard Model and
its Minimal Supersymmetric extension. As a explicit application, we
have considered the impact of these effects on the perturbative
unification of the couplings.

To begin with, we have examined and specified renormalization
procedures for computing the threshold functions. In dealing with
gauge coupling renormalization, we can distinguish two classes of
contributions: those coming from gauge boson vacuum polarization,
which are well and uniquely defined, and contributions coming from
vertex and external legs, which not so clearly defined. In order to
get the latter contribution, we have required that the procedure
respects the Slavnov--Taylor identities. This will guarantee the
universality of the renormalized coupling and, at the same time, will
constrain to some extent the class of particle masses that can
contribute to the vertex function $\Gamma$: only gauge boson
masses. All other massive matter fields contribute to the effective
couplings only through the transverse component of the vacuum
polarization.

For the QCD gauge coupling, since gluons are massless, the vertex
function is simply given by the standard term calculated using an
MISP. For the $SU(2)_L \times U(1)_Y$ gauge couplings, the symmetry is
broken at low energies, and we have massive gauge bosons. To get the
explicit expressions for the vertex function, we follow the arguments
given in Ref. \cite{kl}. In this way, at the same time we get
universal $and$ mass dependent couplings, so that in a single stroke
we solve another problem related to the electroweak symmetry breaking:
the misdiagonalization of the neutral mass matrix.

We remark that we follow Ref. \cite{kl} $only$ in order to define the
vertex function $\Gamma$. Our effective couplings differ from their
definitions. They define the effective couplings including only vector
boson self--energies, that contain at least, one exactly--conserved
matter current from the unbroken subgroup $SU(3)_c \times
U(1)_{em}$. This condition imposes severe constraints on the kind of
matter fields appearing in the effective couplings \cite{lynn}: there
are no contribution from neutral and colorless matter
fields. Therefore, any relation of the couplings relevant for gauge
coupling unification will be independent, for example, of the neutral
higgses and higgsinos. When we study unification with this choice of
the effective couplings, we are interpolating between the unbroken
$SU(3)_c \times U(1)_{em}$ and the unification group $G$, but
forgetting about the electroweak symmetry breaking in the matter
sector of the model.

One can understand the definition of the $g'(q^2)$ and $g(q^2)$ given
in Ref.  \cite{kl} as a direct splitting of the electromagnetic
coupling $e(q^2)$, such as:
\begin{eqnarray}
\frac{1}{e^2(q^2)} &=& \frac{1}{e^2}+\Pi'_{AA}(q^2) \nonumber \\ &=&
                   \left( \frac{1}{g'^2}+\frac{1}{g^2} \right) +
                   (\Pi'_{BA}(q^2)+\Pi'_{W^3A}(q^2) ) \nonumber \\ &=&
                   \frac{1}{g'^2(q^2)}+\frac{1}{g^2(q^2)} \,,
                   \label{rela}
\end{eqnarray}
where the $\Pi'_{ii}$'s include the transverse vacuum polarization and
the appropriate vertex function. We choose, instead, to define
independently the couplings $e^2(q^2)$ and $g^2(q^2)$. Thus, the
relation (\ref{rela}) only fixes the coupling $g'^2(q^2)$. With our
choice, $all$ the particles that interact under $SU(2)_L$ are present,
and affect the evolution of the effective couplings. Of course, the
high energy limit of the effective coupling is the same for both
definitions, as it is dictated by the RGE and the decoupling theorem.

In some sense, our approach is more closely related to the work of
Ref.  \cite{grinstein} on light threshold effects. These authors
related the corrections ($\delta g_i^2$) to the low energy gauge
couplings due to light thresholds (new physics different from the
Standard Model) to experimental quantities measured in scattering
processes. Thus, they define independently $\delta g^2$ and $\delta
(g'^2+g^2)$, the first being related to the transverse vacuum
polarization of the $W^{\pm}$, as we have done.

Our work agree with the one presented in Ref. \cite{grinstein} on the
importance of light and heavy thresholds for the predictions derived
from perturbative unification. In spite of this, the philosophy of our
approach is in some sense different. Taking as unifying group the
supersymmetric extension of $SU(5)$, they extrapolate the relation
given at the unification scale, down to the low energy scale \mz\ as
\begin{equation}
12 \alp{2}-5 \alp{1}-7\alp{3} \mid_{\mz}=0 \,; \label{rela2}
\end{equation}
We only get this relation when using a {\it mass independent}
renormalization procedure. The experimental values of the couplings
are extracted on the bases of the validity of the Standard Model, but
this relationship is obtained when we assume an extended
model. Therefore, if one wants to make use of the experimental data on
the couplings, it is necessary to include the corrections due to the
new degrees of freedom\footnote{We would like to rectify the comment
on the results of Ref. \cite{grinstein} that we included in the {\it
Note added in proof} of Ref. \cite{mia2}. There, we misunderstood the
notion of ``vertex correction" used in Ref. \cite{grinstein}. Of
course, the vertex correction due to new physics (susy masses) are not
universal, and do not have to be included, as they correctly did not
\cite{carta}.}.   This is also the main issue of some recent work
\cite{bagger}. In these papers, a calculation of complete susy
contributions is carried out in order to determine the value of
$\sin^2 \theta_W(\mz)\mid_{\overline{MS}}$ from experimental
quantities, but in the context of the MSSM. These threshold
contributions evaluated $at$ the scale \mz\ tend to decrease the value
of $\sin^2 \theta_W(\mz)$, and as a consequence there is an increase
in the predicted value of \alz\ from unification, hardly compatible
with the experimental one.  In addition to this, we were concerned on
how thresholds effects modified the $evolution$ with scale of gauge
couplings, to understand how in going from $M_X$ to \mz\ these effects
will affect the relation (\ref{rela2}).  We have considered a
more accurate description of these effects, beyond the leading
logarithmic correction given by the step--function approximation.
When including the complete threshold function, not only susy
thresholds have to be considered, but also any threshold effects due
to the massive standard particles. For example, the standard--gauge
boson threshold effects, which become relevant in realizing or not a
perturbative scenario for the unification of the couplings, as we have
shown.

We have found that the predictions derived by imposing perturbative
unification of gauge couplings in the MSSM depend on the procedure
chosen to treat the thresholds. For example, the effective couplings
always favor higher values of \alzi\ and susy masses, than the
calculation with the step--function.  Moreover, although both
procedures give good results at one--loop order without including
heavy thresholds, when we improve the calculation (2--loop order or
include heavy thresholds) the unification with effective couplings
{\it is in conflict} with at $least$ one of the constraints we impose
on the model: the experimental values for \alz\ and the proton
life--time, or the theoretical naturalness bound for susy masses. In
order to maintain the experimental constraints, we would need susy
masses beyond the upper bound of $1\,TeV$ usually required because of
naturalness reasons.

 At least, this is the general conclusion within the framework of
$SU(5)$ unification, where heavy threshold corrections to the
predicted value of \alz\ are positive. One way to avoid this result
would be to considere a different unification group, which gives the
reverse sign for this contribution \cite{bagger}. We would have to
look for a cancellation between ``light'' and ``heavy'' threshold
effects.

\newpage
\appendix
\section{}
In this Appendix we give the analytical expressions for the functions
$\Pi_{ii}^T(q^2)$ and $\Gamma (q^2)$ introduced in Section
\ref{sect2}. These expressions have been calculated using dimensional
regularization, and can be written as combinations of the following
integrals:
\widetext
\FL
\begin{eqnarray}
B_0(a_1,a_2) &=&\int_0^1\!\! dx \ln (a_1x+a_2(1-x)+x(1-x)) \,,\\
B_3(a_1,a_2)&=&\int_0^1\!\! dx x(1-x)\ln (a_1x+a_2(1-x)+x(1-x)) \,,\\
C_0(a_1,a_2,\xi a_2)&=&\int_0^1\!\! dx \int_0^x \!\!dy \ln
(a_1(1-x)+a_2(x-y+\xi y)+x(1-x)) \,,\\
D_1(a_1,a_2,\xi a_2)&=&\int_0^1\!\! dx \int_0^x \!\!dy \frac{1-x}
{(a_1(1-x)+a_2(x-y+\xi y)+x(1-x))} \,,\\
D_3(a_1,a_2,\xi a_2)&=&\int_0^1\!\! dx \int_0^x \!\!dy \frac{x(1-x)}
{(a_1(1-x)+a_2(x-y+\xi y)+x(1-x))} \,,\\
(1-\xi )H(a_1,a_2)&=& \nonumber \\ \!&\displaystyle{\frac{1}{2
a_2}}&\!\int_0^1\!\! dx \int_0^x \!\!dy \ln \frac{a_1(1-x+\xi
(x-y))+a_2 y+y(1-y)}{a_1(1-x+\xi (x-y))+\xi a_2 y+y(1-y)}\,,
\end{eqnarray}
\narrowtext
\noindent and the linear combinations:
\begin{equation}
D_{31}^+=D_1+D_3\;\;,\;\;\;\;D_{31}^-=D_1-D_3\,.
\end{equation}
Here, we have introduced the variable
$\displaystyle{a_i=\frac{m^2_i}{-q^2}}$, where ($-q^2$) is the
euclidean momentum.

The general transverse contributions to the $\Pi_{ii}^T$, which depend
on the kind of masses running in the loops, are given by ($\xi$ is the
gauge parameter):

(a) Gauge Boson + Ghost:
\begin{eqnarray}
\lefteqn{(4 \pi)^2 \Pi^{(\sg)}(a_1,a_2)=-\frac{1}{2}\Biggl\{ \left(
\frac{13}{3}-\xi \right) \left( \frac{2}{\varepsilon}-\ln
\frac{-q^2}{\mu^2} \right) + \frac{2}{3} } \nonumber \\ & & -2
B_0(a_1,a_2)-10 B_3(a_1,a_2)+2 B_3(\xi a_1,\xi a_2) \nonumber \\ & &
-(1-\xi ) \biggl( 1+C_0(a_1,a_2,\xi a_2)+C_0(a_2,a_1,\xi a_1)
\nonumber \\ & & +(1+a_1) D_{13}^+(a_1,a_2,\xi a_2)+(1+a_2)
D_{13}^+(a_2,a_2,\xi a_2) \biggr) \nonumber \\ & & +(1-\xi)^2
H(a_1,a_2) \Biggr\} \,.
\end{eqnarray}

(b) Scalar--Gauge Boson: \widetext
\begin{equation}
(4 \pi)^2 \Pi^{(\ss \sg)}(a_1,a_2)= -(1-\xi) a_2 D_{13}^-(a_1,a_2,\xi
a_2)\,,
\end{equation}
\narrowtext
\noindent Here $m^2_1$ is the scalar mass, and $m^2_2$ the gauge boson
mass.

(c) Scalars: \widetext
\begin{equation}
(4 \pi)^2 \Pi^{(\ss)}(a_1,a_2)=\frac{1}{3} \left\{ \left(
\frac{2}{\varepsilon}-\ln \frac{-q^2}{\mu^2} \right) -3 B_0(a_1,a_2) +
12 B_3(a_1,a_2) \right \} \,. \label{pis}
\end{equation}
\narrowtext

(d) Fermions:
\widetext
\begin{equation}
(4 \pi)^2 \Pi^{(\sf)}(a_1,a_2)=\frac{4}{3} \left\{ \left(
\frac{2}{\varepsilon}-\ln \frac{-q^2}{\mu^2} \right) -6 B_3(a_1,a_2)
\right \} \,.
\label{pif}
\end{equation}
\narrowtext

In the expressions above, we have not included the couplings or group
factors.  The scale``$\mu$" is the unit of mass introduced in
dimensional regularization, and $\varepsilon$ is given, as it is usual
in Modified Minimal Subtraction, by
$\displaystyle{\frac{2}{\varepsilon}=\frac{2}{n-4}-\gamma+\ln 4 \pi}$.

In each particular case, we get: \widetext
\begin{eqnarray}
\lefteqn{\Pi_{AA}^T(q^2)= e^2 \biggl\{ 2 \Pi^{(\sg)}(a_{\sw},a_{\sw})
+\Pi^{(\ss \sg)}(\xi a_{\sw},a_{\sw})+ \Pi^{(\ss)}(\xi a_{\sw},\xi
a_{\sw}) } \nonumber \\ & & +\sum _f Q_f^2 \Pi^{(\sf)}(a_f,a_f)
\biggr\} \,, \label{piaa} \\
\lefteqn{\Pi_{WW}^T(q^2)= g^2 \biggl\{ 2 s_{\sw}^2
\Pi^{(\sg)}(0,a_{\sw})+ 2 c_{\sw}^2 \Pi^{(\sg)}(a_{\sz},a_{\sw})
}\nonumber \\ & &+\Pi^{(\ss \sg)}(a_h,a_{\sw})+s_{\sw}^2 \Pi^{(\ss
\sg)}(\xi a_{\sw},0) +s_{\sw}^4 \Pi^{(\ss \sg)}(\xi a_{\sw},a_{\sz})
\nonumber \\ & &+\frac{1}{4} \Pi^{(\ss)}(a_h,\xi a_{\sw}) +\frac{1}{4}
\Pi^{(\ss)}(\xi a_{\sz},\xi a_{\sw}) + \frac{1}{4}\sum
_{doublets}\Pi^{(\sf)}(a_{f_1},a_{f_2}) \biggr\} \,,
\label{piww} \\
\lefteqn{\Pi_{gg}^T(q^2) = g_3^2 \biggl\{ 3 \Pi^{(\sg)}(0,0)
+\frac{1}{2} \sum _{quarks} \Pi^{(\sf)}(a_f,a_f) \biggr\}
}\,. \label{pigg}
\end{eqnarray}
\narrowtext

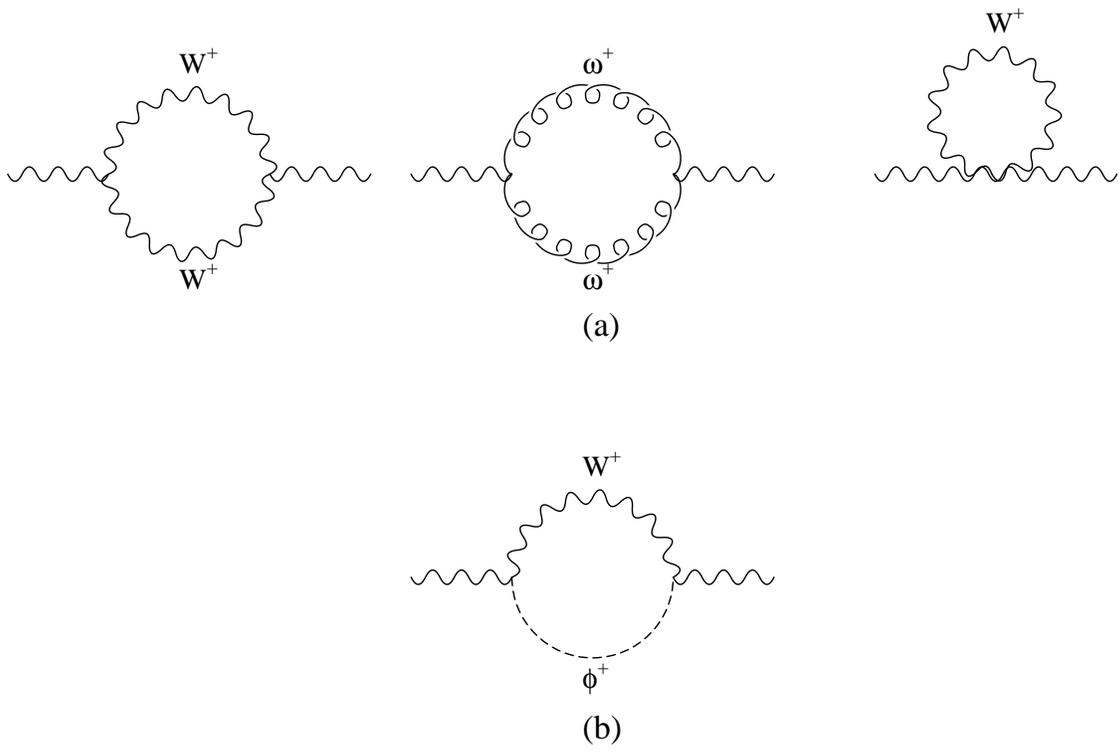
\begin{figure}\caption{ The 1--loop particle contributions to the $\Pi^L_{ZA}$
function, $i.e.$ to $\Gamma^{'}$. (a) Gauge boson plus ghost. (b) Scalars and
gauge bosons. \label{figgam}}
\end{figure}

In order to obtain the function $\Gamma (q^2)=-(g^2+g'^2)
\Pi^L_{ZA}(q^2)/(g g') m^2_{\sz}$, we need the longitudinal term of
the diagrams in Fig.  (\ref{figgam}). The diagrams (\ref{figgam}.a)
and (\ref{figgam}.b) are related and we do not need to calculate both
explicitly; the same kind of diagrams contribute to the longitudinal
vacuum polarization of the photon, which we know to be
zero. Therefore, if we name $A(a_{\sw})$ the longitudinal term of
(\ref{figgam}.a), and $B(a_{\sw})$ the corresponding to
(\ref{figgam}.b), they must verify the following identity, \widetext
\begin{equation}
\Pi_{AA}^L(q^2)= e^2 m_{\sw}^2 \left( A(a_{\sw})+B(a_{\sw}) \right)=0
\rightarrow A(a_{\sw})=-B(a_{\sw}) \,.
\end{equation}
So that, for $\Pi_{ZA}^L(q^2)$ we obtain:
\begin{equation}
\Pi_{ZA}^L(q^2)=-e \frac{g^2}{\sqrt{g^2+g'^2}} m_{\sw}^2 A(a_{\sw}) +
e g \frac{g'^2}{g^2+g'^2}m_{\sz} m_{\sw} B(a_{\sw})= g g'm_{\sw}^2
B(a_{\sw})\,,
\end{equation}
where,
\begin{eqnarray}
B(a_{\sw})&=&\biggl\{ \frac{3+\xi}{2}\left( \frac{2}{\varepsilon}-\ln
\frac{-q^2}{\mu^2} \right) -2 B_0(\xi a_{\sw},a_{\sw}) \nonumber \\ &
&+(1-\xi) C_0(\xi a_{\sw},a_{\sw},\xi a_{\sw}) -2 (1-\xi) D_{13}^-(\xi
a_{\sw},a_{\sw},\xi a_{\sw}) \biggr\} \,,
\end{eqnarray}
\narrowtext
\noindent And finally, we get for $\Gamma(q^2)$:
\begin{equation}
\Gamma(q^2)=(g^2+g'^2) \frac{m_{\sw}^2}{m_{\sz}^2}B(a_{\sw})= -g^2
B(a_{\sw})\,.
\label{defgam}
\end{equation}

As we did in the main body of the paper, on the following we redefine
$\Pi^T_{ii} \equiv g_i^2 \Pi^T_{ii}$ and $\Gamma \equiv g^2 \Gamma$.

Now, we write the expressions for $\Pi^T_{ii}(q^2)+2 \Gamma_i(q^2)$ in
the Landau gauge, and separate in each term the divergent part:
\begin{eqnarray}
\lefteqn{(4 \pi)^2 (\Pi_{AA}^T+ 2 \Gamma (q^2))= } \nonumber \\ &
&-\frac{13}{3}
\left(\frac{2}{\varepsilon}-\ln\frac{-q^2}{\mu^2}+F_{\sg}(a_{\sw},a_{\sw})
\right)
-3\left(\frac{2}{\varepsilon}-\ln\frac{-q^2}{\mu^2}+F_{\sgam}(a_{\sw})
\right) - F_{\ss \sg}(0,a_{\sw}) \nonumber \\ & &+\frac{1}{3}
\left(\frac{2}{\varepsilon}-\ln\frac{-q^2}{\mu^2} +F_{\ss}(0,0)
\right) +\frac{4}{3}\sum_f Q_f^2
\left(\frac{2}{\varepsilon}-\ln\frac{-q^2}{\mu^2}+F_{\sf}(a_f,a_f)
\right) \,,\label{pigame}\\
\lefteqn{(4 \pi)^2 (\Pi_{WW}^T+ 2 \Gamma (q^2))=} \nonumber \\ & &
-\frac{13}{3} \left(\frac{2}{\varepsilon}-\ln\frac{-q^2}{\mu^2}+
s_{\sw}^2 F_{\sg}(0,a_{\sw})+ c_{\sw}^2 F_{\sg}(a_{\sz},a_{\sw})
\right)
-3\left(\frac{2}{\varepsilon}-\ln\frac{-q^2}{\mu^2}+F_{\sgam}(a_{\sw})
\right) \nonumber \\ & &- F_{\ss \sg}(a_h,a_{\sw})-s_{\sw}^2 F_{\ss
\sg}(0,0) -s_{\sw}^4 F_{\ss \sg}(a_{\sz},0)
+\frac{1}{3}\sum_{doublets}
\left(\frac{2}{\varepsilon}-\ln\frac{-q^2}{\mu^2}+F_{\sf}(a_{f_1},a_{f_2})
\right) \nonumber \\ & & +\frac{1}{6}
\left(\frac{2}{\varepsilon}-\ln\frac{-q^2}{\mu^2}+\frac{1}{2}F_{\ss}(a_h,0)
+\frac{1}{2}F_{\ss}(0,0)\right) \,, \label{pigam2}\\
\lefteqn{(4 \pi)^2 (\Pi_{gg}^T+ 2 \Gamma_3 (q^2)) = } \nonumber \\ &
 &-11 \left(\frac{2}{\varepsilon}-\ln\frac{-q^2}{\mu^2}+Cte \right)
 +\frac{2}{3}\sum_{quarks}
 \left(\frac{2}{\varepsilon}-\ln\frac{-q^2}{\mu^2}+F_{\sf}(a_f,a_f)
 \right) \,.
\label{pigam3}
\end{eqnarray}
The functions $F_j(a_i)$ contain the threshold effects associated to
massive degrees of freedom. Each of these functions has the property
that they tend to a constant in the limit where the masses can be
neglected as compared to the scale, and they behave as $\ln a_i +
O(1/a_i)$ in the limit of heavy masses compared with the momentum
scale; except the function $F_{\ss \sg}(a_i,a_j)$, which tends to zero
in both limits. Having in mind these limits, we can study the behavior
of the effective couplings, $$
\frac{1}{g_i^2(q^2)}=\frac{1}{g_i^2(q_0^2)}+\left( \Pi_i^T(p^2)+2
\Gamma_i(p^2) \right)\bigg|_{p^2=q_0^2}^{p^2=q^2}\,,
$$
for several ranges of scale:

(a) Limit $q^2 > q_0^2 \gg m_i^2$: \widetext
\begin{eqnarray}
\frac{(4 \pi)^2}{e^2(q^2)}&=& \frac{(4 \pi)^2}{e^2(q_0^2)}+ \left(
-\frac{13}{3}+\frac{4}{3} \sum_f Q_f^2\right) \ln \frac{q_0^2}{q^2} =
\frac{(4 \pi)^2}{e^2(q_0^2)}+\frac{11}{3} \ln \frac{q_0^2}{q^2} \,, \\
\frac{(4 \pi)^2}{g^2(q^2)}&=&\frac{(4 \pi)^2}{g^2(q_0^2)}+ \left(
-\frac{13}{3}+\frac{1}{6}+\frac{1}{3} \sum_{doublets}\right) \ln
\frac{q_0^2}{q^2} = \frac{(4 \pi)^2}{g^2(q_0^2)}-\frac{19}{6} \ln
\frac{q_0^2}{q^2} \,, \\
\frac{(4 \pi)^2}{g_3^2(q^2)}&=& \frac{(4 \pi)^2}{g_3^2(q_0^2)}+ \left(
-11+\frac{2}{3} \sum_{quarks}\right) \ln \frac{q_0^2}{q^2} = \frac{(4
\pi)^2}{g_3^2(q_0^2)}-7 \ln \frac{q_0^2}{q^2} \,.
\end{eqnarray}
\narrowtext
\noindent Thus, we recover in this limit (negligible masses) the usual
expressions derived by minimal subtraction.

(b) Limit $q_0^2 < q^2 \ll m_i^2$: in this case, the massive degrees
of freedom must decouple and not contribute to the effective
couplings.  When working in the Landau gauge, we have to pay special
attention to the contribution of the Goldstone bosons and ghosts. In
the Landau gauge, we treat these degrees of freedom as massless, and
we could argue that the limit of heavy gauge masses will not affect
them. If we proceed in this way, we will get the non--decoupling
effects proportional to the gauge masses. The correct result is
obtained reverting the limits: first, we take the limit $m_i^2/q^2
\rightarrow \infty$ in a general gauge, and then we take $\xi
\rightarrow 0$. We will assume this procedure whenever we take the
limit of heavy masses\footnote{For example, we have considered this
when we talk before about the limits of the function $F_{\ss
\sg}(a_i,a_j)$.}.

(c) Limit $q_0^2 \ll m_i^2 \ll q^2$: In this case we recover, in part,
the results obtained by using the step--function approximation. For
example, for the electromagnetic coupling we get:
\begin{eqnarray}
\frac{(4 \pi)^2}{e^2(q^2)}&=&\frac{(4 \pi)^2}{e^2(q_0^2)}
+\lim_{m^2/q^2 \rightarrow 0} \left( \Pi_i^T(q^2)+2 \Gamma_i(q^2)
\right) -\lim_{m^2/q_0^2 \rightarrow \infty} \left( \Pi_i^T(q_0^2)+2
\Gamma_i(q_0^2) \right) \nonumber \\ &=&\frac{(4
\pi)^2}{e^2(q_0^2)}-7\left(
\ln\frac{m_{\sw}^2}{q^2}+\frac{17}{14}-\frac{2}{21} \right)
+\frac{4}{3}\sum_f Q_f^2 \left( \ln\frac{m_f^2}{q^2}+\frac{5}{3}
\right) \nonumber \\ &=&\frac{(4 \pi)^2}{e^2(q_0^2)}-7\ln
c_{\sw}\frac{m_{\sw}^2}{q^2}+\frac{2}{3} +\frac{4}{3}\sum_f Q_f^2 \ln
c_f \frac{m_f^2}{q^2} \,.
\end{eqnarray}
In the last line, we simply have absorbed the constant terms in the
logarithms, except for the factor ``2/3", which has a different
origin\footnote{It is a typical threshold effect of the massive gauge
bosons, produced by the regularization method employed. Indeed, it is
not present when using dimensional reduction \cite{dr}.}. The
differences between this expression and the one obtained by the
step--function method are in these constants. Their origin is better
understood if we study the $\beta$ function directly, which is given
by: \widetext
\begin{eqnarray}
(4 \pi)^2 \frac{\beta_e(q^2)}{e^4(q^2)}&=& -\frac{d}{d \ln q^2} \left(
\Pi_i^T(q^2)+2 \Gamma_i(q^2) \right) \nonumber \\ &=& -\frac{13}{3}
f_g(a_{\sw})-3 f_{\sgam}(a_{\sw})
+\frac{1}{3}f_s(0,0)+\frac{4}{3}\sum_f Q_f^2 f_f(a_f) \,,
\end{eqnarray}
where we have defined,
\begin{eqnarray}
& & \frac{13}{3} f_g(a_{\sw})= \frac{13}{3}\left( 1-\frac{d
F_{\sg}(a_{\sw},a_{\sw})}{d \ln q^2}\right)- \frac{d F_{\ss
\sg}(0,a_{\sw})} {d \ln q^2} \,, \\ & &f_{\sgam}(a_{\sw})= 1- \frac{d
F_{\sgam}(a_{\sw})}{d \ln q^2}\,, \\ & &f_f(a_f)=1- \frac{d
F_{\sf}(a_f,a_f)} {d \ln q^2} \;,\;\;\;\; f_s(0,0)=1-\frac{d
F_{\ss}(0,0)} {d \ln q^2} \,.
\end{eqnarray}
\narrowtext
\noindent We can seen in Fig. (\ref{derivf2}), where we have plotted
$f_f(a_f)$, that the behavior of the functions $f_i(a_j)$ is similar
to the step--function.  When we integrate $\beta_e$ from $q_0^2$ to
$q^2$, we can divide the integration range into $[q_0^2,m_i^2]$ and
$[m_i^2,q^2]$, and approximate the functions $f_i(a_i)$ in each
interval for the corresponding limit ($m_i^2/q_0^2 \rightarrow
\infty$, $m_i^2/q^2 \rightarrow 0$). In this way, we reproduce exactly
the results of the step--function approximation. Nevertheless, if
instead we use the Taylor expansion of the functions $f_i(a_j)$ to
first order in $a_i$ (improving the latter approximation), we would
get for a fermion of mass $m_f$, \widetext
\begin{eqnarray}
- \int_{q_0^2}^{q^2} f_f(m^2_f/y) d\ln y &\simeq & -
\int_{1}^{m_f^2/q_0^2} \frac{1}{c_f^{\infty} y} d\ln y +
\int_{1}^{m_f^2/q^2} (1+c_f^0 y) d\ln y \nonumber \\ &=&
\frac{q_0^2}{c_f^{\infty} m^2_f} -\frac{1}{c_f^{\infty}}+\ln
\frac{m_f^2}{q^2}+ c_f^0 \frac{m^2_f}{q^2}- c_f^0 = \ln c
\frac{m_f^2}{q^2} \,,
\end{eqnarray}
\narrowtext
\noindent where once again a constant appears in the logarithm. But,
neither of the two approximations, step--function or Taylor expansion,
is valid, because when we integrate we cross the region $m_f^2/q^2
\approx 1$, where none of these approximations is defined. Due to
this, the best procedure is to approximate the complete function,
$f_f(a_f)$, by an expression valid for the full energy range, as it is
\cite{georgi}
\begin{equation}
f_f(a_f) \simeq \frac{1}{1+c_f a_f} \,.
\end{equation}
When we integrate with this function, we get: \widetext
\begin{equation}
- \int_{q_0^2}^{q^2} f_f(m^2_f/y) d\ln y \simeq
-\int_{m^2_f/q^2}^{m^2_f/q_0^2} \frac{1}{1+c_f y} d \ln y = ln
\frac{q_0^2 +c_f m^2_f}{q^2 +c_f m^2_f}\,.
\end{equation}
\narrowtext
\noindent Therefore, we have seen that, for any massive particle, its
contribution to the effective coupling can be approximated by a
logarithmic function of the form:
\begin{equation}
L_k(q^2,m^2)=ln \frac{q_0^2 +c_k m^2}{q^2 +c_k m^2}\,.
\label{approx}
\end{equation}
With this function, we recover easily the limits (a) and (b) for the
effective couplings and, on the other hand, we get a more precise
behavior of the thresholds for intermediate scales. The value of the
constant $c_k$ depends on the kind of particles we treat (fermion,
scalar,..), and if it is or not degenerated with its partner in the
loop. In particular for the Standard Model, with or without
supersymmetry, and the range of energies of interest, the
contributions from loops of non--degenerate particles can be treated
as completely degenerate, or that one of the masses can be neglected
with respect to the other. A numerical study shows that the best
values of the constant $c_k$ for these cases, are those given in Table
\ref{consc}.

In the following, and to close this Appendix, we include the
analytical expressions of the functions $F_j(a_1,a_2)$:
\begin{eqnarray}
\lefteqn{F_{\sf}(a_1,a_2)=\frac{5}{3}-\frac{1}{2}\ln a_1 a_2
-2(a_1+a_2)-2(a_1-a_2)^2} \nonumber \\ & &-\frac{1}{2}(a_1-a_2)
(3(a_1+a_2)+2(a_1-a_2)^2) \ln\frac{a_1}{a_2} \nonumber \\ &
&+\frac{1}{2}(1-a_1-a_2-2(a_1-a_2)^2) R_{12} L_{12} \,, \\
\lefteqn{F_{\ss}(a_1,a_2)=\frac{8}{3}-\frac{1}{2}\ln a_1 a_2
+4(a_1+a_2)+4(a_1-a_2)^2} \nonumber \\ & &+\frac{1}{2}(a_1-a_2)
(3+6(a_1+a_2)+4(a_1-a_2)^2) \ln\frac{a_1}{a_2} \nonumber \\ &
&+\frac{1}{2}(1+2(a_1+a_2)+4(a_1-a_2)^2) R_{12} L_{12} \,, \\
\lefteqn{F_{\ss \sg}(a_1,a_2)=\frac{1}{6}\bigl( a_2-2 a_2^2+4 a_2 a_1
} \nonumber \\ & & +(1+3 a_1+3a_1(a_1-a_2)+(a_1-a_2)^3)\ln
\frac{a_2}{a_1}\nonumber \\ & & -(1+2 a_1-a_2+(a_1-a_2)^2)
R_{12}L_{12} -2 (1+a_1)^3 \ln \frac{1+a_1}{a_1} \bigr)\,,\\
\lefteqn{F_{\sgam}(a)=\frac{5}{3}-\ln a+\frac{a}{3}-\frac{1}{3 a}\ln
(1+a) -(1+a+\frac{a^2}{3})\ln \frac{1+a}{a} } \\
\lefteqn{\frac{3}{13} F_{\sg}(a_1,a_2)=A(a_1,a_2)+a_1 B(a_1,a_2)+ a_2
B(a_2,a_1)} \nonumber \\ &
&+\frac{C(a_1,a_2)}{a_1}+\frac{C(a_2,a_1)}{a_2}+\frac{a_2}{a_1}
D(a_1,a_2) +\frac{a_1}{a_2} D(a_2,a_1)+\frac{E(a_1,a_2)}{a_1 a_2} \,,
\end{eqnarray}
\begin{eqnarray}
A(a_1,a_2)&=&\frac{121}{18}-\frac{7}{3}\ln a_1 a_2+6 a_1 a_2
+\frac{13}{3}a_1a_2(a_2-a_1)\ln\frac{a_2}{a_1}\nonumber \\ &
&+(\frac{19}{12}+3a_1 a_2)R_{12}L_{12} \,, \label{aa}\\
B(a_1,a_2)&=&
-4-\frac{8}{3}a_1+(\frac{3}{8}-\frac{17}{6}a_1-\frac{7}{6}a_1^2)
\ln\frac{a_1}{a_2} -(\frac{13}{12}+\frac{4}{3}a_1)R_{12}L_{12} \,, \\
C(a_1,a_2)&=&-\frac{7}{24}\ln a_1 a_2 +\frac{7}{12}\ln
(1+a_2)+\frac{1}{3}R_{12} L_{12} \,, \\
D(a_1,a_2)&=&\frac{1}{24} \biggl( (23+17a_2-2 a_2^2-4a_2^3)(2 \ln
(1+a_2)-\ln a_1 a_2) \nonumber \\ & &+ (15+2a_2-4a_2^2)R_{12}L_{12}
\biggr) \,, \\
E(a_1,a_2)&=&\frac{1}{24} \left( \ln a_1 a_2 -2\ln (1+a_1)-2\ln
(1+a_2)-R_{12}L_{12} \right) \,,
\end{eqnarray}
where we have introduced:
\begin{equation}
\left\{ \begin{array}{l} R_{12}= \sqrt{1+2 (a_1+a_2)+(a_1-a_2)^2} \,,
\\ L_{12}=\ln \left|
\displaystyle{\frac{1+a_1+a_2-R_{12}}{1+a_1+a_2+R_{12}}} \right|
\,. \end{array} \right.
\end{equation}

We note that as regularization procedure we use dimensional
regularization to work with the Standard Model, and dimensional
reduction to work within the MSSM. Contributions from fermion and
scalars are the same in both procedures, the only difference being in
the gauge boson contribution. For translating to dimensional
reduction, we have only to subtract a constant term ``$2/3$'' from
equation (\ref{aa}).

\section{}
In this appendix we approximate the effective couplings at the 2--loop
order. The RGE for gauge couplings at the 2--loop order are given, in
general, by \cite{2loop,marciano}
\begin{equation}
\frac{d \alpha_i(\mu)}{d \ln \mu}=\frac{b_i(\mu)}{2 \pi}
\alpha_i^2+\frac{b_{ij}(\mu)}{8 \pi^2} \alpha_j(\mu) \alpha_i^2(\mu)
\,.
\label{rge2b}
\end{equation}
When we calculate the coefficient $b_i$ and $b_{ij}$ using an MISP, we
obtain constant values for $b_i$ and $b_{ij}$; but when using an MDSP
and we include the dependence on the masses, these coefficients gain a
dependence with the scale $\mu$, through the ratios $(m_k/\mu)$. The
$b_i(\mu)$ take into account the threshold effects at 1--loop order,
that we have calculated in Appendix A. Now, we want evaluate which is
the threshold contribution at 2--loop order, included in the
$b_{ij}(\mu)$, with respect to that at 1--loop order, without actually
performed the explicit calculation.

When we integrate (\ref{rge2b}) from \mz\ to an arbitrary scale $\mu$,
we get: \widetext
\begin{equation}
\alpha_i^{-1}(\mu)=\alpha_i^{-1}(\mu)\mid_{1-loop} - \frac{1}{4 \pi}
\int_{m_{\sz}}^{\mu} \frac{b_{ij}(\mu')}{b_j(\mu')} d\ln
\alpha_j(\mu') \,.
\end{equation}
\narrowtext
\noindent In order to evaluate the contribution of the second term, we
choose an intermediate scale, $\mz<\mu_1<\mu$, such that $m_i \ll
\mu_1$, for all the masses $m_i$. For example, for the Standard Model
and its supersymmetric extension, we have $m_i \leq 1\,TeV$, and thus
it is sufficient if we take $\mu_1=10\,TeV$. In the range $\mu'\geq
\mu_1$, the thresholds at 1--loop order and 2--loop order are
negligible, and then the coefficients $b_i(\mu')$ and $b_{ij}(\mu')$
remain as constants. Therefore, we can write:
\begin{eqnarray}
\alpha_i^{-1}(\mu)\!&=&\!\alpha_i^{-1}(\mu)\mid_{1-loop} - \frac{1}{4
 \pi} \int_{m_{\sz}}^{\mu_1} \frac{b_{ij}(\mu')}{b_j(\mu')} d\ln
 \alpha_j(\mu') - \frac{1}{4 \pi} \int_{\mu_1}^{\mu}
 \frac{b_{ij}}{b_j} d\ln \alpha_j(\mu') \nonumber \\ \!&=&\!
 \alpha_i^{-1}(\mu)\mid_{1-loop} - \frac{1}{4 \pi}
 \int_{m_{\sz}}^{\mu} \frac{b_{ij}}{b_j} d\ln \alpha_j(\mu') -
 \frac{1}{4 \pi} \int_{m_{\sz}}^{\mu_1}
 \left(\frac{b_{ij}(\mu')}{b_j(\mu')} -\frac{b_{ij}}{b_j}
 \right)d\ln\alpha_j(\mu') \nonumber \\ \!& \equiv &\!
 \alpha_i^{-1}(\mu)\mid_{1-loop} +\frac{b_{ij}}{4 \pi b_j} \ln
 \frac{\alpha_j(m_{\sz})}{\alpha_j(\mu)}+ \sum_j \Delta_{ij}\,.
\end{eqnarray}
Now, the 2--loop threshold effects are included in the function
$\Delta_{ij}$, and this is the function in which we are interested. To
evaluate the order of magnitude of this function, we approximate the
integrand in $\Delta_{ij}$ by a straight line in the variable $\ln
\alpha_j(\mu')$,
\begin{equation}
\left(\frac{b_{ij}(\mu')}{b_j(\mu')} -\frac{b_{ij}}{b_j} \right)
\simeq a_{ij}+c_{ij} \ln \alpha_j(\mu') \,,
\end{equation}
with the condition that it passes through the points \mz\ and $\mu_1$,
i.e., \widetext
\begin{eqnarray}
\mu'=m_{\sz} &\rightarrow& a_{ij}+c_{ij} \ln \alpha_j(m_{\sz}) =
\left(\frac{b_{ij}(m_{\sz})}{b_j(m_{\sz})} -\frac{b_{ij}}{b_j} \right)
\equiv B_{ij}(m_{\sz}) \,, \\ \mu'=\mu_1 &\rightarrow& a_{ij}+c_{ij}
\ln \alpha_j(\mu_1) = \left(\frac{b_{ij}(\mu_1)}{b_j(\mu_1)}
-\frac{b_{ij}}{b_j} \right) \simeq 0 \,.
\end{eqnarray}
\narrowtext
\noindent With these conditions, the expression for $\Delta_{ij}$ is
reduced to:
\begin{equation}
\Delta_{ij}=\frac{B_{ij}(m_{\sz})}{8 \pi} \ln \frac{\alpha_j(m_{\sz})}
{\alpha_j(\mu)} \,.
\end{equation}
The final step consists of taking at 2--loop order the same kind of
threshold functions, $f_i(m_i/\mu)$, that at 1--loop order, i.e., a
function bounded between 0 and 1. For example, $b_{23}\sim 8+4
f_{top}$, and then the value of $\Delta_{23}$ will be between 0.013
($f_{top}=0$) and 0.003 ($f_{top}=1$).  Checking for different values,
the corrections to $\alpha_i^{-1}$, given by $\Delta_i=\sum_j
\Delta_{ij}$, are never higher than 0.3\%, and therefore, negligible.

Thus, the 2--loop effective couplings can be approximated by:
\begin{equation}
\alpha_i^{-1}(\mu)=\alpha_i^{-1}(\mu)\mid_{1-loop} +\frac{b_{ij}}{4
\pi b_j} \ln \frac{\alpha_j(m_{\sz})}{\alpha_j(\mu)}\,,
\end{equation}
where we only need to include explicitly the threshold functions at
1--loop order.

 This approximation is independent of the chosen regularization
procedure, in particular if dimensional regularization or dimensional
reduction. The regularization procedure does not affect the
coefficients of the RGE's for the gauge coupling when we work with
\msu\, and always preserve its form (at one or two loops). With an
MDSP, it easily seen that the function $b_i(\mu)$ at one --loop order
does not depend on the regulator. When working at two--loop order, we
would have to check how the $explicit$ expressions of the
$b_{ij}(\mu)$--coefficients depend on the regularization
method. However, here we have only made use of the general properties
of these functions, i.e., their behavior in the limiting cases $m/\mu
\rightarrow 0$ and $\mu/m \rightarrow 0$. And this behavior is
independent of the regularization procedure, as well of the particular
mass dependent subtraction procedure one uses.

\newpage
%

\newpage
%
%
\narrowtext
\begin{table}
\caption{Experimental values of $\alpha_3(m_Z)$.}
  \begin{tabular}{|c|c|c|}
\hline
 Experiment & Central Value  & Error  \\
\hline
     ALEPH jets      & 0.125         & $\pm 0.005$    \\
    DELPHI jets     & 0.113         & $\pm 0.007$    \\
    DELPHI ($e^+e^-$)& 0.118         & $\pm 0.005$    \\
    L3 jets         & 0.125         & $\pm 0.009$    \\
    OPAL jets       & 0.122         & $\pm 0.006$    \\
    OPAL $\tau$     & 0.123         & $\pm 0.007$    \\
\hline
    $J/\Psi$         & 0.108         & $\pm 0.005$    \\
   Deep Inelastic   & 0.111         & $\pm 0.005$    \\
   UA6              & 0.112         & $\pm 0.009$    \\
\hline
  \end{tabular}
\label{table3}
 \end{table}
%
\begin{table}
\caption{ Fitted values of $c_k$.}
\begin{tabular}{lc}
Massive particles &
 $c_k$ \\
1 Gauge boson &  2 \\
2 Gauge boson &  5 \\
1 Fermion     & 2.5 \\
2 Fermions    &  5 \\
1 Scalar      &  4 \\
2 Scalars     & 10 \\
 \hline
\end{tabular}
\label{consc}
\end{table}
\widetext
%
%
\begin{table}
\caption{ Values calculated with the step--function approximation$^{\rm a}$.}
  \begin{tabular}{|c|c|c|c|c||c|c|c|c|c|}
\hline
$m_0^{max}$ & $\xi_0^{max}$&$M_{\Phi}^{min}$& $M_{\Sigma}^{max}$
&$\alpha_3^{min}$
&$m_{1/2}^{max}$&$\xi_0^{min}$&$m_0^{min}$
& $M_{\Phi}^{max}$&$\alpha_3^{max}$
 \\
\hline
1000 & 494 & $10^{16.58}$&$10^{14.5}$&0.112&91&121&742&$10^{16.79}$&0.125
\\
2000 &1975 & $10^{16.06}$&$10^{16.3}$&0.108&358&31&742&$10^{16.79}$&0.125
\\
\hline
  \end{tabular}
\label{susypt}
$^{\rm a}$ Mass values in $GeV$.
\end{table}
%
%
\begin{table}
\caption{Values calculated with the effective couplings$^{\rm a}$.}
  \begin{tabular}{|c|c|c|c|c|c||c|c|c|c|}
\hline
$m_0^{max}$ & $\xi_0^{max}$&$m_{\tilde{h}}^{min}$&$M_{\Phi}^{min}$&
$M_{\Sigma}^{max}$&$\alpha_3^{min}$&
$m_{1/2}^{max}$&$\xi_0^{min}$& $M_{\Phi}^{max}$
&$\alpha_3^{max}$
\\
\hline
1536 & 1165&1536 & $10^{16.26}$ & $10^{16.49}$ & 0.125 & 45 &1165
&$10^{16.26}$ & 0.125
\\
3000 &4444 & 370 & $10^{15.75}$ & $10^{16.53}$ & 0.120 & 239 & 158
&$10^{16.50}$ & 0.125
\\
4000 &7901 & 198 & $10^{15.53}$ & $10^{16.54}$ & 0.118 &794&  26
&$10^{16.61}$ & 0.125
\\
5000 &12348& 121 & $10^{15.35}$ & $10^{16.55}$ & 0.116 &2784&   3
&$10^{16.69}$ & 0.125
\\
\hline
  \end{tabular}
\label{susytt}
$^{\rm a}$ Mass values in $GeV$.
\end{table}

\begin{references}
\bibitem[*]{aa1} Also at: Instituto de Matem\'aticas y F\'{\i}sica
Fundamental, C.S.I.C., Serrano 119--123, 28006 Madrid.

\bibitem[**]{aa2} Also at: Instituto de Matem\'aticas y F\'{\i}sica
Fundamental, C.S.I.C., Serrano 119--123, 28006 Madrid and Theoretical
Division, Los Alamos National Laboratory, Los Alamos, New Mexico 87545

\bibitem{lam2} N.--P. Chang, A. Das and J. P\'erez--Mercader,
\plb{93}{1980}{137}; \prd{22}{1980}{1414}.

\bibitem{lam1} S. Weinberg, \plb{91}{1980}{51}; C. H. Llewelyn Smith, G.
G. Ross and J. F. Wheather, \npb{177}{1981}{263}; L. Hall,
\npb{178}{1981}{75}; P. Bin\'etruy and T. Sch\"uker,
\npb{178}{1981}{307}.

\bibitem{msm} W. A. Bardeen, A. J. Buras, D. W. Duke and T. Muta,
\prd{18}{1978}{3998}.

\bibitem{misp} S. Weinberg, \prd{8}{1973}{3497}; J. C. Collins and A. J.
Macfarlane, \prd{10}{1974}{1201}.

\bibitem{mdsp} H. Georgi and H. D. Politzer, \prd{14}{1976}{1829}; W. Celmaster
and R. J. Gonsalves, \prd{20}{1979}{1420}; O. Nachtmann and W. Wetzel,
\npb{146}{1979}{272}.

\bibitem{desac} T. Appelquist and J. Carazzone, \prd{11}{1975}{2856}.

\bibitem{marciano} W. J. Marciano and A. Sirlin, in
{\it Proc. of the Second Workshop on Grand Unification},
Ann Arbor, 1981, ed. J. Leveille, L. Sulak and D. G. (Birkh\"ausser, Boston).

\bibitem{amaldi} M. B. Einhorn and D. R. T. Jones, \npb{196}{1982}{475};
W. J. Marciano and G. Senjanovi\'c, \prd{25}{1982}{3092};
J. Ellis, S. Kelley and D. V. Nanopoulos, \plb{249}{1990}{441};
U. Amaldi, W. de Boer and H. Furstenau, \plb{260}{1991}{447};
P. Langacker and M. Luo, \prd{44}{1991}{817}.

\bibitem{zichichi} F. Anselmo, L. Cifarelli,
A.  Peterman and A. Zichichi, \nca{105}{1992}{1025}; V. Barger,
M. S. Berger and P.  Ohmann, \prd{47}{1993}{1093}; P. Langacker and
N. Polonsky, \prd{47}{1993}{4028}; M. Carena, S. Pokorski and
C. E. M. Wagner, \npb{406}{1993}{59}.

\bibitem{gell} M. Gell--Mann and F. E. Low, \pr{95}{1954}{1300};
 E. C. G. Stuckelberg and A. Peterman, {\sl Helv. Phys. Acta} {\bf 26}
(1953) 499.

\bibitem{lynn} B. W. Lynn, preprint SU--ITP--93--22 (1993).

\bibitem{grinstein} A. Faraggi and B. Grinstein, \npb{422}{1994}{3}.

\bibitem{bagger} J. Bagger, K. Matchev and D. Pierce,
preprint JHU-TIPAC-95001 (1995);
P. H. Chankowski, Z. Pluciennik and S. Pokorski, preprint IFT-94/19 (1994).

\bibitem{ross} D. A. Ross, \npb{140}{1978}{1}; T. Goldman and D. A. Ross,
\npb{171}{1980}{272}.

\bibitem{bc3} R. Coqueraux, \prd{23}{1981}{1365}.

\bibitem{zg} D. J. Gross and F. Wilczek, \prd{8}{1973}{3633}.

\bibitem{kl} D. C. Kennedy and B. W. Lynn, \npb{322}{1989}{1}; D. C. Kennedy,
\npb{351}{1991}{81}.

\bibitem{sirlin} W. J. Marciano and A. Sirlin, prd{22}{1980}{2695}.

\bibitem{kuroda} M. Kuroda, B. Moultaka and D. Schildknecht,
\npb{350}{1991}{73}.

\bibitem{mia1} M. Bastero--Gil, V. Man\'{\i}as and J. P\'erez--Mercader,
Int. Journal of Mod. Phys. A 10 (1995) 373.

\bibitem{data} {\it Review of Particle Properties}, \prd{50}{1994}{1173}.

\bibitem{top} CDF Collab., F. Abe et. al, \prd{50}{1994}{2966}.

\bibitem{mh} J. L. Lopez, D. V. Nanopoulos, H. Pois, X. Wang and A. Zichichi,
\plb{306}{1993}{73}.

\bibitem{maiani} N. Cabbibo, L. Maiani, G. Parisi and R. Petronzio,
\npb{158}{1979}{295}; R. A. Flores and M. Sher, \prd{27}{1983}{1679};
M. Lindner, \zpc{31}{1986}{295}.

\bibitem{ale} G. Degrassi, S. Fanchiotti and A. Sirlin, \npb{351}{1991}{49}; H.
Burkhardt, F. Jegerlehner, G. Penso and C. Verzegnassi,
\zpc{11}{1989}{497}.


\bibitem{nath} P. Nath and R. Arnowitt, \prl{69}{1992}{725};
\plb{287}{1992}{89}; J. L. Lopez, D. V. Nanopoulos, H. Pois and A. Zichichi,
\plb{299}{1993}{262}.

\bibitem{hisano} J. Hisano, H. Murayama and T. Yanagida, \prl{69}{1992}{1014};
\npb{402}{1993}{46}.

\bibitem{lsp} J. L. Lopez, D. V. Nanopoulos and A. Zichichi,
\plb{291}{1992}{255}; J. L. Lopez, D. V. Nanopoulos and H. Pois,
\prd{47}{1993}{2478}; P. Nath and R. Arnowitt, \plb{299}{1993}{58}; $ibid$. B
303 (1993) 403 (E); \prl{70}{1993}{3696}; R. G. Roberts and L.
Roszkowski, \plb{309}{1993}{329}.

\bibitem{mia2} M. Bastero--Gil and J. P\'erez--Mercader, \plb{322}{1994}{355}.

\bibitem{lahanas} A. B. Lahanas and D. V. Nanopoulos, Phys. Rep. 145 (1987) 1.

\bibitem{dis} W. Wong, P. B. Mackenzic, R. Rosenfeld and J. L. Rosner,
\prd{37}{1988}{3210}; A. D. Martin, R. G. Roberts and W. J. Stirling,
Durham Preprint DTP 90--76 (1990); \plb{266}{1991}{173};
\prd{47}{1993}{867}.

\bibitem{lepqcd1} ALEPH Collab., D. Decamp et. al., \plb{284}{1992}{163};
DELPHI Collab., P. Abreu et. al., \zpc{54}{1992}{55}; L3 Collab.,
O. Adriani et. al., \plb{284}{1992}{471}; OPAL Collab., P. D. Acton,
\zpc{55}{1992}{1}.

\bibitem{lepqcd2} UA6 Collab., G. Ballocchi et. al., \plb{317}{1993}{250};
ALEPH Collab., D. Decamp et. al., \plb{307}{1993}{209}; DELPHI
Collab., P.  Abreu et. al., \plb{311}{1993}{408}.

\bibitem{pd5} S. Weinberg, \prd{26}{1982}{287}; N. Sakai and T. Yanagida,
\npb{197}{1982}{533}; S. Dimopoulos, S. Raby and F. Wilczek,
\plb{112}{1982}{133}; B. A. Campbell, J. Ellis and D. V. Nanopoulos,
\plb{141}{1984}{221}; R. Arnowitt, A. H. Chamseddine and P. Nath,
\plb{156}{1985}{215}; \prd{32}{1985}{2348}; \prd{38}{1988}{1479}.

\bibitem{carta} We thank B. Grinstein for constructive correspondence on this
matter.

\bibitem{dr} W. Siegel, \plb{84}{1979}{193}; D. M. Cooper, D. R. T. Jones and
P. van Nieuwenhuizen, \npb{167}{1980}{479}.

\bibitem{georgi} H. Georgi {\it et. al} in \cite{mdsp}

\bibitem{2loop} D. R. T. Jones, \prd{25}{1982}{581}; M. B. Eihorn and D. R. T.
Jones, \npb{196}{1982}{475}.

\end{references}
\end{document}